\pdfoutput=1

\documentclass[acmsmall, screen]{acmart}
\setcopyright{none}
\pdfoutput=1
\usepackage{wrapfig}
\usepackage{colortbl}
\usepackage{tabularray}
\usepackage{color}
\usepackage[normalem]{ulem}
\usepackage{adjustbox}

\usepackage{amsthm} 

\usepackage{color,soul}
\usepackage{graphics}
\usepackage{hyperref}
\usepackage{url}
\usepackage{lipsum}
\usepackage{tikz}
\usepackage{enumitem}	
\usepackage{listings} 
\usepackage{booktabs}	
\usepackage{lipsum}
\usepackage{subcaption}
\usepackage{capt-of}	
\usepackage{diagbox}
\usepackage{multirow}
\usepackage{todonotes}  
\definecolor{refkey}{rgb}{249,158,26}
\definecolor{labelkey}{rgb}{0,0,1}
\definecolor{airforceblue}{rgb}{0.36, 0.54, 0.66}
\definecolor{applegreen}{rgb}{0.55, 0.71, 0.0}

\usepackage{mathrsfs}	
\usepackage{amsfonts}
\usepackage{algorithm,amsmath,algpseudocode} 
\usepackage{boldline}					

\usepackage{comment}
\usepackage{multirow}

\definecolor{frenzyorange}{RGB}{249, 158, 26}

\renewcommand{\paragraph}[1]{\vskip 3pt\noindent\textbf{#1 }}	 

  {\begin{list}{$\bullet$}%
     {\setlength{\parsep}{0pt}%
      \setlength{\topsep}{0pt}%
      \setlength{\itemsep}{2pt}}}%
  {\end{list}}
\newcommand\Noted[1]{} 

\definecolor{darkblue}{rgb}{0.0, 0.0, 0.55}
\definecolor{mygreen}{HTML}{ADFF2F}
\definecolor{mylightgray}{gray}{0.8}

  {\begin{itemize}
	[leftmargin=0cm,
		itemindent=.3cm,
		labelwidth=\itemindent,
		labelsep=0pt,
		parsep=1pt,
		topsep=1pt,
		itemsep=1pt,
		align=left]
  }%
  {\end{itemize}}    

  {\begin{enumerate}
	[leftmargin=.cm,itemindent=.5cm,labelwidth=\itemindent,
		labelsep=0pt,
		parsep=1pt,
		topsep=1pt,
		itemsep=3pt,
		align=left]
  }%
  {\end{enumerate}}    

\newcommand{\mTwoUltra}{M2 Ultra}

\makeatletter
\def\@copyrightspace{\relax}
\makeatother

\setcopyright{acmlicensed}
\copyrightyear{2025}
\acmYear{2025}
\acmDOI{XXXXXXX.XXXXXXX}

\acmJournal{JACM}
\acmVolume{37}
\acmNumber{4}
\acmArticle{111}
\acmMonth{8}

\usepackage{xcolor}      
\usepackage{mdframed}    
\newmdenv[
  backgroundcolor   = white,      
  linecolor         = gray,       
  linewidth         = 1pt,        
  roundcorner       = 3pt,        
  innertopmargin    = 6pt,
  innerbottommargin = 6pt,
  innerleftmargin   = 8pt,
  innerrightmargin  = 8pt,
  skipabove         = 1em,
  skipbelow         = 1em,
]{takeawaybox}

\usepackage[skins]{tcolorbox}
\tcbuselibrary{skins,breakable}
\usepackage{hyperref}

\newtcolorbox{block}[2][]{%
  enhanced,
  breakable,
  fonttitle = \bfseries,
  coltitle     = black,
  title        = #2,
  sharp corners, left=4pt,right=4pt,top=2pt,bottom=2pt,
  #1
}

\begin{document}



\title{Profiling Large Language Model Inference on Apple Silicon: A Quantization Perspective}


\author{Afsara Benazir}
\email{hys4qm@virginia.edu}
\affiliation{%
  \institution{University of Virginia}
  \city{Charlottesville}
  \state{Virginia}
  \country{USA}
}

\author{Felix Xiaozhu Lin}
\affiliation{%
  \institution{University of Virginia}
  \city{Charlottesville}
  \state{Virginia}
  \country{USA}
}

\email{felixlin@virginia.edu}
\begin{abstract} 

 
 
A systematic understanding of Apple Silicon is lacking in the current landscape of hardware efficiency; research focus is largely centered on accelerating GPUs for large-scale training or inference on CUDA devices. 
This paper investigates Apple Silicon's unique memory architecture that offers a unified memory integrating CPU and GPU memory and its implications for on-device LLM inference.  

We decipher myths about whether Apple Silicon is efficient for on-device inference compared to competitors such as NVIDIA GPUs by directly conducting latency and throughput comparison benchmarks. We explain the performance gap between them through profiling low level hardware metrics - ALU utilization, memory bandwidth, buffer usage, cache residency etc. at runtime. We draw several insights regarding performance bottlenecks such as dequantization overhead, compute throughput and memory bandwidth.
We debunk existing false claims regarding large language model inference such as compressing models to lower bit precision is a defacto promise for faster inference across all hardware platforms. We find that the large unified memory enables Apple Silicon to be both cost effective and efficient against NVIDIA GPUs for ultra large language models.

Our work provides a
comprehensive perspective of on-device inference on commodity GPUs; our large scale evaluation on 5 hardware testbeds incorporating three Apple M-series devices: M2 Ultra, M2 Max and M4 Pro and two NVIDIA GPUs: NVIDIA RTX A6000, a multi GPU setup with 2xNVIDIA RTX A6000, 5 model scales ranging from 8B to 405B parameters and 14 quantization schemes gives an understanding of how Apple Silicon fits within the paradigm of on-device LLM inference. Our analysis reveals multiple resource interdependencies and unexpected findings, while also quantifying established insights. 
To the best of our knowledge, this study makes the first attempt to present a thorough characterization and analysis of Apple Silicon for on-device inference.
\end{abstract}

\maketitle

\section{Introduction}

The emergence of billion scaled large language models (LLMs) and their higher reasoning capabilities has lead to a profound paradigm shift in the growth of artificial intelligence \cite{minaee2024large}.
On-device execution of large LLMs is lucrative as it promises enhanced privacy, data localization, user control of data and allows for versatile personalized applications \cite{xu2024device}. With the recent advancement in machine learning (ML) accelerators, GPUs can provide peak performance;
a limiting factor in accessing the supreme capabilities of these GPUs are their increasingly high cost that can be measured in terms of \$\$ per token generated \cite{shekhar2024towards}. Although cloud providers offer on-demand GPU instances, purchasing GPUs are more cost-effective in the long run \cite{strubell2020energy}.


A significant constraint for LLM inference is the substantial demand for GPU memory - loading tens of GB of static model parameters alongside storing the intermediate activations and KV cache in limited GPU memory is inefficient. Devices like Apple Mac Studio \cite{m2_ultra} can support upto 192 GB of unified memory making it a prime candidate for hosting LLMs. To date, 22.9 million units of Apple Silicon chips are in circulation \cite{mac-sale-stat}, which is on the rise. 
Although NVIDIA GPUs are dominating the market with their higher compute utilization and tensor cores dedicated for batch processing in LLM training and finetuning \cite{cheng2024thorough}, their contemporary Apple Silicon (ref. \S\ref{sec:as}) is overlooked owing to its reported lower GFLOP count \cite{hubner2025apple}, despite its availability of a large unified memory. 

For a private LLM serving workstation intended for personal use or by a small user cohort, two factors are critical - (1) inference speed - the number of tokens generated per second and (2) cost effectiveness: how much money's worth one can get out of a single inference run, measured through \$/token.

For LLM training and finetuning, tokens are processed in batches, where GPU shines owing to its supreme parallel processing power. 
LLM inference is autoregressive - during \textit{prefill phase}, all input prompt tokens are processed in parallel to generate the very first token, but in the \textit{decode phase} 
each subsequent token is generated conditioned on the previous one.
When hosting an LLM engine that serves a number of parallel requests such as in a cloud server, batched processing can be done but for single request model serving, typical for on-device inference scenarios, decoding is truly sequential. 

A higher parallel processing capability (i.e higher throughput) alone does not ensure faster token generation speed as (1) for a single request LLM inference, the decode phase is memory bound \cite{kim2023squeezellm}, efficient compute units are left idle as data transfer happens (2) For inference of extremely/ultra-large language models in the range of 70B-405B parameters, a challenge is to fit the model in the GPU VRAM to be efficiently processed; a limitation for current consumer grade GPUs having a limited 24GB-48 GB VRAM \cite{song2024powerinfer}. 

One solution is to lower the memory traffic during each decoding pass, thereby boosting generation speed. 
Weights in LLM models are typically represented in high precision formats (FP16/FP32), representing these weights using fewer bits sharply reduce that traffic. Post-training quantization (PTQ) \cite{yao2024exploring} shines here owing to its capability to produce coherent output at extremely low bits per weight. But translating the memory reduction achieved through quantization to latency improvement can be challenging as it requires significant engineering efforts \cite{jin2024comprehensive}.


\paragraph{\textbf{Motivation}} Our motivation stems from the unified memory offered by Apple Silicon at lower cost per million tokens generated (ref. \S\ref{sec:rc}) which is attractive for on-device inference. The bigger memory pool allows for effective data exchange, dynamic allocation and reduced overhead (no data duplication); even if raw compute power is lower \cite{kenyon2022apple}, it avoids the performance penalties associated with slow offloading to system RAM that discrete GPUs with comparatively lower VRAM might encounter. This calls for a closer examination of Apple Silicon’s strengths and weaknesses, and evaluating its potential as a strong competitor for on-device ML inference.

\textbf{Goal} A lack of systematic understanding of LLM inference on Apple Silicon exists - prior work has primarily focused on LLM training \cite{cheng2024thorough} and on-device LLM inference on NVIDIA GPUs\cite{li2024large, lin2405qserve, luo2024benchmarking, song2024powerinfer} or ARM CPUs \cite{gope2024highly}. To address this gap in literature we conduct a principled empirical study of Apple M-series devices under 26 different model precisions for absolute runtime latency, cost-effectiveness and to understand the impact of hardware characteristics on efficiency.
Our observations are reported as a series of "findings" for ML practitioners to make informed choices in LLM serving.
We identify the following important research questions (RQ):
\begin{itemize}


\item RQ1: How well does Apple Silicon support end-to-end LLM inference on-device?

\item RQ2: Is Apple Silicon a cost effective choice for on-device LLM inference compared to other hardware accelerators such as CUDA?

\item RQ3: What are the real hardware bottlenecks for on-device inference in Apple Silicon — hardware compute capability, DRAM bandwidth limitation or de-quantization overhead? 
\end{itemize}

This paper is the first to investigate Apple's unique memory architecture and its implications for on-device LLM inference. 
We conduct a top-down quantitative evaluation of multiple Apple Silicon generations using various quantization schemes to characterize end-to-end performance and address RQ1 (\S\ref{sec:e2e}).  
To address RQ2, we compare against two NVIDIA device configurations, evaluating both single-GPU and multi-GPU setups to assess cost-effectiveness (\S\ref{sec:comparison}).
We dissect the inference run of models across different bit precisions to pinpoint the performance bottleneck, resource utilization and answer RQ3 (\S\ref{sec:eb}).
\paragraph{\textbf{Finding Summary}}
Running ultra-large models such as Llama 405B with
acceptable latency is impossible on contemporary consumer GPUs (NVIDIA, AMD etc.) but is possible on Apple Silicon as it offers a single
machine workstation without the hassle of configuring a heterogeneous device. Through several micro-benchmarks we  empirically validate that on Apple Silicon, model size is not proportional to faster inference - a 2-bit model can be faster than a 1-bit one (ref. \S\ref{block:2}). Apple Silicon becomes increasingly cost effective against contemporary GPUs as the model parameter scales (ref. \S\ref{block:3}). We find that a lack of dedicated compute units (similar to tensor cores in CUDA) on Apple Silicon holds back its performance (ref. \S\ref{block:9}). 
Additionally, codebook-based quantization schemes significantly slows down token generation speed on Apple Silicon, underscoring the need to align quantization design with hardware architecture (ref. \S\ref{sec:pc}).
At low bit precision, the dequantization overhead is significant, Apple Silicon becomes bound by arithmetic operations more than the memory bandwidth (ref. \S\ref{block:7}).


\paragraph{\textbf{Contribution}}
In this work (1) we try to understand the performance capabilities, overhead and limitations of Apple Silicon during LLM inference.
(2) Perform detailed profiling and analysis of inference runtime to comprehend the impact of varied model precision on Apple Silicon.
(3) We empirically validate that (a) lower bit precision doesn’t guarantee faster inference; 
latency is primarily determined by the underlying hardware characteristics and runtime bottlenecks (b) Apple Silicon is both a cost‑effective and efficient choice for on‑device LLM inference, particularly for ultra‑large models.
(4) We recommend and demonstrate that for Apple silicon, block based quantization schemes instead of codebook based schemes are the superior choice for deployment. 


\textbf{Roadmap} The remainder of this paper is organized as follows: \autoref{sec:background} discusses the preliminary background of this work. 
\autoref{sec:methodolody} outlines our methodology, detailing the hardware and software setup, choice of models used in evaluation and characterization of our chosen evaluation metrics.
\autoref{sec:rc} details the overall inference cost in terms of latency and cost per million tokens-both per stage and end-to-end and reports a comparison benchmark between Apple Silicon and CUDA devices.
\autoref{sec:eb} delves into kernel and hardware-level utilization of compute and memory units and analyzes bottlenecks to interpret the findings from \autoref{sec:rc}. 
In \autoref{sec:recommendation}, we provide recommendation for ML practitioners and hardware vendors to optimize LLM serving on Apple Silicon.




\section{Background} 
\label{sec:background}
\subsection{Apple Silicon} 
\label{sec:as}
\begin{wrapfigure}{r}{0.35\textwidth}  
  \centering
    \includegraphics[width=0.3\textwidth]{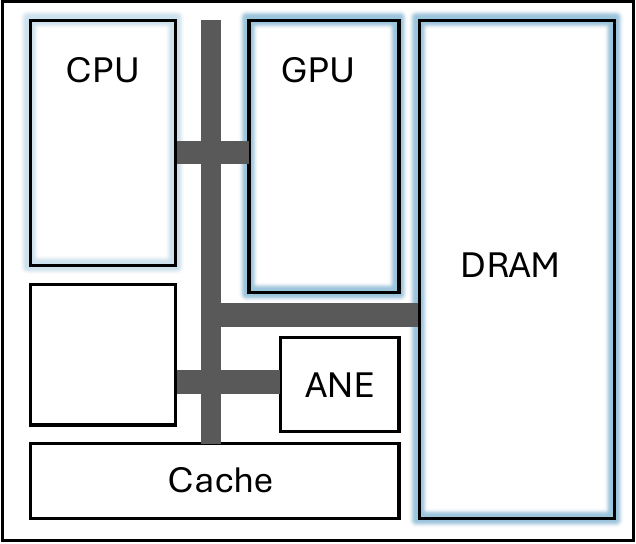}
  \caption{Schematic design of Apple M-series System-on-Chip (SoC)}
  \label{fig:apple_soc}
\end{wrapfigure}

Apple introduced its propreitary M-series System-on-Chips (SoCs) in 2020 \cite{apple-silicon-wiki} that champions a Unified Memory Architecture (UMA) where the CPU, GPU and Apple Neural Engine (ANE) are tightly integrated and share one large memory pool, reducing data movement overhead between them (ref. \autoref{fig:apple_soc}). 
In contrast NVIDIA's architecture consists of discrete GPUs having VRAM separate from the host RAM \cite{luo2024benchmarking}; a design engineered to maximize raw computational throughput, especially for massive parallel workloads needed for large scale LLM training. A fundamental trade-off thus emerges - the benefits of integrated and shared-resource efficiency (in Apple) versus the potential of peak compute performance with dedicated GPU workstations. 

\textit{Terminology}: Apple's primary API for GPU programming is Metal \cite{metal}, which provides low-level access to its hardware; for NVIDIA the interface is CUDA \cite{ghorpade2012gpgpu}. ANE is Apple's neural processing unit (NPU) with accelerated compute but has support for only small machine learning models.



\subsection{On-device LLM inference}

Billion scaled LLMs trained on trillions of tokens are capable of delivering highly human-like interactions. Owing to their enormous size, such large LMs are typically hosted on cloud servers where users can access it through a paid API. Running these models on consumer-grade GPUs is impractical unless they are compressed but this compression comes at the loss of accuracy. Quantization can drastically reduce model size and latency by leveraging faster arithmetic \cite{jin2024comprehensive}.

GPUs thrive at parallel processing, ideal for large scale LLM training or finetuning done in batches.
But inference typically happens for a few parallel requests, often for a single batch/request. During inference, owing to auto-regressive decoding the bottleneck shifts from compute to memory as the device sits idle waiting for data to be transferred from the host \cite{williams2009roofline}. If the GPU memory is not sufficient to hold all of the model parameters, certain layers are offloaded to CPU \cite{sheng2023flexgen} - this introduces additional communication overhead \cite{cheng2024thorough}.
Research on on-device LLM inference has focused so far on CUDA GPUs \cite{luo2024benchmarking, kwon2023efficient, recasens2025mind} or ARM CPUs \cite{gope2024highly}. While some efforts have been made to characterize Apple GPU performance \cite{hubner2025apple}, they are not specifically targeted towards on-device 
LLM use cases.


\subsection{Quantization}
Post-training quantization (PTQ) \cite{jin2024comprehensive} is a popular technique for lossless compression of model weights to lower bits per weight (bpw), lucrative for on-device inference.
Quantizing a full precision FP32 Llama 405B model at 1.6 TB of memory to 2 bits reduces model size by 11x. 
Symmetric quantization maps real values to uniformly spaced levels using a constant scale factor 
$\Delta$  with each value quantized via 
q=round(x/$\Delta$) and clamped to a fixed integer range \cite{gholami2022survey}. This linear, round-to-nearest (RTN) approach underpins many PTQ methods and is the defacto standard for most hardware accelerators e.g. \textbf{legacy} 4-bit, 8-bit etc. 
Asymmetric quantization introduces a zero-point offset to allocate finer granularity to more important weights.
\cite{gholami2022survey, kim2023squeezellm}. This non-uniform mapping through logarithmic scales or K-means clustering \cite{dettmers2023spqr} better preserves important outlier values.

\textbf{Block quantization} as exemplified by GPTQ \cite{frantar2022gptq}, QLora \cite{dettmers2023qlora}, K-quants of llama.cpp \cite{llama_cpp} etc. divides weight matrices into fixed-size blocks called superblocks (e.g. size 32, 64, or 256) with a scale and offset; further divided into subblocks (consisting of 8 or 16 weights) having its own independent scale and offset. Furthermore, each layer of a transformer \cite{vaswani2017attention} block uses mixed schemes, quantizing critical layers at higher precision, leaving non-critical layers at lower precision.
\textbf{Vector quantization (VQ)} \cite{gong2025survey} or codebook based quantization represents the vector of multiple elements within a weight tensor as a single element, that serves as the index of a custom codebook such as in QUIP\# \cite{tseng2024quip}, GPTVQ \cite{van2024gptvq}, AQLM \cite{egiazarian2024extreme}, IQ quants in llama.cpp \cite{llama_cpp}. 
The codebook can be crafted using K means clustering or using a lattice (Hessian mattrix, E8 lattice etc.) and its size can impact latency \cite{egiazarian2024extreme}.  To prevent referencing a large codebook, llama.cpp employs 
a 3rd-order polynomial to map codebook entries e.g. IQ4\_NL uses a 16-entry 8-bit integer codebook with a non-uniform  NF \cite{dettmers2023qlora} like distribution to map 4-bit quantized indices into 8-bit integer value.

At runtime, the quantized integers are \textit{dequantized} by approximating floating-point values using either the block-specific scale and offset or in the case of codebook-based quantization, the corresponding codebook entries.




\vspace{-5pt}
\subsection{Roofline model}
The roofline model \cite{williams2009roofline} compares compute operations against hardware boundaries. It shows which resource limits the application’s throughput: if memory bound - the kernel moves more data (larger weights); if compute-bound, it spends more time on arithmetic operations (inside ALUs). Arithmetic intensity (AI) quantifies how much compute is done per byte of data moved 
and is a typical metric for quantifying this behavior,  a high AI means the kernel is compute bound and vice versa. 
In the context of single batch inference, the \textit{memory wall} problem \cite{kim2023squeezellm} that shows the imbalance between compute and memory boundness, is extremely challenging to address. Typically, prefill is compute‑bound as tokens are processed in parallel; decode is memory bound \cite{verma2023mastering} - the high dimensional weight matrices loaded into registers is utilized only once for a single decode token, keeping the compute units underutilized. Despite the availability of high bandwidth DRAM in recent times, it still cannot match the high compute power of GPU cores provided by vendors \cite{gholami2022survey, qureshi2020tearing}, that are intended for large scale training in mind.


\vspace{-5pt}
\section{Methodology}
\label{sec:methodolody}
\begin{table}[t]
\centering
\includegraphics[width=\linewidth]{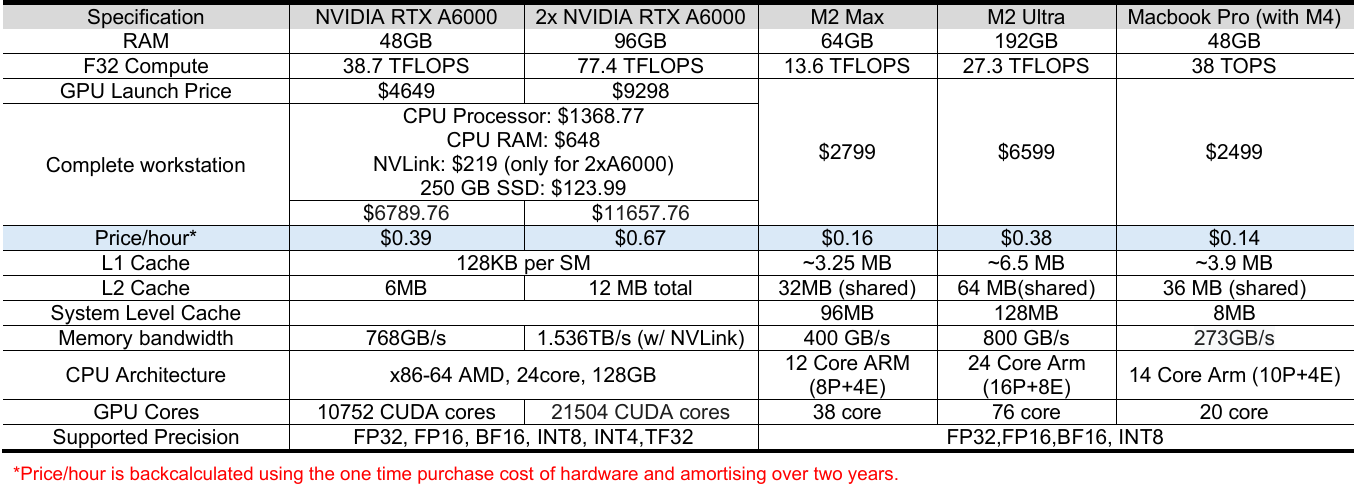}
\caption{Device specification.}
\label{tab:device}
\vspace{-10pt}
\end{table}

\setlength{\textfloatsep}{5pt}
\textbf{Hardware Platform}
Our experimental hardware is illustrated in \autoref{tab:device}  and consists of four hardware testbeds and 5 device configurations: an M2 Max equipped with 64 GB of unified memory, an M2 Ultra with 192 GB, M4 Pro with 48 GB, a single RTX A6000 featuring 48 GB and a dual-RTX A6000 setup with a combined 96 GB VRAM. We focus exclusively on commodity GPUs serving personal use workstations that serve single request inference and do not consider data center GPUs such a H200/A100 which is extremely expensive and intended for model training/finetuning.

\textit{Apple M-series GPU}: Mac Studio equipped with M2 Ultra is a small form-factor workstation from Apple with 24 CPU Cores (16 performance cores and 8 efficiency cores), 76 GPU cores containing {9,728} ALU units and 192 GB of unified memory. It corresponds to 2x M2 Max chips glued together, each M2 Max having 13.6 TFLOPS of FP32 performance; so a total of 27.2 TFLOPS in M2 Ultra \cite{m2_ultra}. Its 24-core CPU has a measured integer math capability of approximately 117.5 GOps/s. Approximately 75\% of the available RAM is actively usable \cite{apple-tech-talk}. 
The M2 Max’s 512-bit memory bus delivers up to 400 GB/s of bandwidth, and the M2 Ultra simply doubles that to around 800 GB/s. The Macbook Pro with M4 chip represents the latest generation of Apple SoCs offering 273 GB/s memory bandwidth.


\textit{NVIDIA CUDA}: NVIDIA RTX A6000 is an Ampere-based workstation GPU with 10,752 CUDA cores (equivalent to ALU units) or 84 SMs (streaming multiprocessors)  delivering up to 38.7 TFLOPS of single precision performance and 309.7 TFLOPS of dedicated FP16 tensor compute \cite{nvidia-wiki}, for demanding ML workload.
A single A6000 provides 768 GB/s memory bandwidth; a dual‑A6000 setup sums to about 1.5 TB/s local bandwidth, while the NVLink \cite{nvlink} between cards is 112.5GB/s.
The CPU used is the AMD Ryzen Threadripper 3960X (24‑core/48‑thread). 



\textbf{Estimation of price/hour} or 
\$\$/GPU\_hour is back-calculated by taking the hardware’s upfront purchase price and
amortizing it over a two-year period.
Based on their launch price \cite{mac-launch-price}, price per hour for M2 Max, M2 Ultra and M4 Pro is \$0.159, \$0.376 and \$0.143 respectively.
As NVIDIA GPUs do not come with an integrated workstation, a complete setup increases its configuration price thus increasing its price per hour to \$0.384 (1×A6000) and \$0.665 (2×A6000). The price for the complete workstation is derived in \autoref{tab:device} combining GPU pricing data \cite{nvidia-launch-price} with remaining components - CPU processor: \$1368.77 \cite{amd-ryzen-cpu}, 128 GB host memory: \$648 \cite{cpu-mem-price}, 1 TB PCIe 3.0x4 + NVMe SSD: \$123.99 \cite{ssd-price} and NVLink: \$219 \cite{nvlink-price}. Our approximation of the complete workstation is humble as the cloud rate for a single A6000 is \$0.49/hour \cite{runpod} compared to ours estimated at \$0.39/hour.

\textbf{Software setup}
We use Apple Instruments and Apple Xcode v16.4 - two development toolkit from Apple that can capture a GPU trace and profile performance counters. For our LLM serving framework we employ llama.cpp \cite{llama_cpp} [build d6d2c2ab]. The end to end latency measurements are done using this framework and fine grained utilization, limiter, miss rate etc. values are obtained from the profiling tools. 

\begin{table}[t]
    \centering \includegraphics[width=0.85\linewidth]{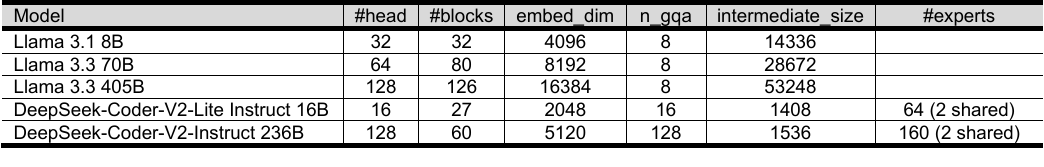}
    \vspace{-5pt}
    \caption{Architecture of models used in evaluation.}
    \label{fig:arch}
    \vspace{-5pt}
\end{table}
\setlength{\textfloatsep}{5pt}
\textbf{Model Setup}
Our model zoo includes five selections illustrated in \autoref{fig:arch}: Llama 3.1 8B, Llama 3.3 70B, and Llama 3.3 405B from the Llama model family \cite{grattafiori2024llama}; DeepSeek-Coder-V2-Lite Instruct 16B and DeepSeek-Coder-V2-Instruct 236B from the DeepSeek-V2 \cite{liu2024deepseek} line. Llama uses the decoder-only transformer architecture while Deepseek models are predominantly of mixture-of-experts (MoE) architecture which activate only a subset of experts at runtime, reducing VRAM overhead. Both model family offer advanced reasoning capabilities and high performance and they each come in multiple parameter sizes; we chose these five to showcase a wide span of model scales. The models are downloaded from huggingface in GGUF format \cite{gguf}, a binary file format that is optimized for fast loading and saving. Moving further we refer to the models as Llama 8B, Llama 70B, Llama 405B, Deepseek 16B and Deepseek 236B.




\textbf{Quantization Scheme} We analyze 26 distinct quantization variants implemented in llama.cpp, covering both block-based (K-quants) and codebook-based (IQ-quants) schemes. Bits per weight (bpw) quantifies the average number of bits used to represent each weight in a quantized model. In the K-quants naming convention, the prefix "Q" followed by a digit (e.g. Q2\_K) indicates the bit-width used for quantization (e.g., 2.62 bits), with "\_K" indicating strictly block-wise quantization. IQ quants (e.g. IQ1\_M) can be row wise or block wise but the "IQ" prefix signifies codebook-based quantization, with the digit indicating bit-width (1.75 bits). Suffixes S (small), M (medium) or L (large) denote a choice of mixed precision configuration that results in a fraction of bits less or more per weight (e.g., IQ1\_M stores output projections in Q6\_K). 



\textbf{Evaluation Metric}
\label{sec:em}
We choose from a broad spectrum of evaluation metrics:

(1) \textbf{Tokens per second (TPS)} as calculated by the number of tokens generated per second during a single-request inference. A complement to this is the \textbf{per token latency} which is the time taken to generate one token (during prefill or decode, lower the better). We alternate between the two as a metric of runtime throughput.

(2) \textbf{Cost per million tokens} is represented as the amount in dollars required to generate one million tokens (lower the better). Comparing the cost of CUDA and Metal GPUs can be nuanced, as Metal GPUs come integrated within a complete Mac desktop system that includes storage and other components, offering a more comprehensive, ready-to-use package - whereas NVIDIA GPUs are typically sold as standalone units without storage. As a rough approximation we report cost per million tokens in correspondence with \cite{erdil2025inference}. 
$$\$\$/{1M\ tokens} = \frac{\$/GPU\_hour * 10^{6}}{TPSx3600}$$

We compare our choice of evaluation metric with respect to (1) model runtime: prefill or decode and (2) hardware features: memory usage, operator granularity, performance limiters, overhead and arithmetic intensity in order to get a holistic understanding of our LLM workload across different bit precision on Apple Silicon.


\section{Runtime Cost}
\label{sec:rc}
We attempt to answer RQ1 and RQ2 through this empirical study and primarily look into the end to end inference latency on Apple Silicon across several quantization schemes and as secondary compare it to NVIDIA GPUs.

\begin{figure}
    \centering
    \includegraphics[width=0.55\linewidth]{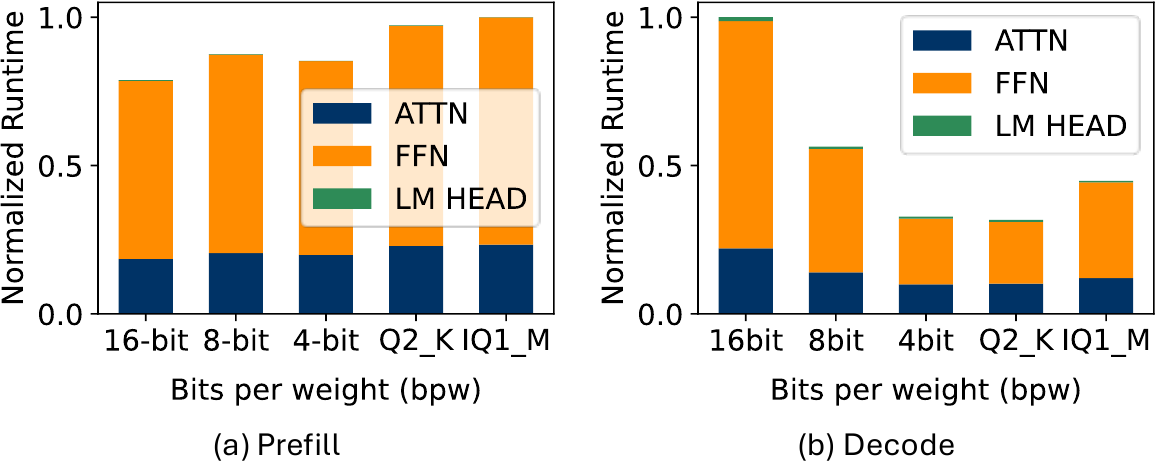}
    \vspace{-5pt}
    \caption{Normalized runtime of Llama 70B on M2 Ultra.}
    \label{fig:nr}
\end{figure}

\subsection{Per stage runtime cost}
\autoref{fig:nr} illustrates the normalized runtime of primary blocks of a transformer model during prefill and decode for an end to end inference.
We measure the time distribution of the major execution blocks (attention: ATTN, feed-forward: FFN, output projection: LM HEAD) the ratio of which remains constant across different bit widths. FFN takes up 76\% of the time for a single token generated in a 16 bit model.

\begin{block}{}
    \textbf{Finding \#1 (a)}: Both in prefill and decode stage of dense models, operations in feed-forward layers are the most expensive taking up 76\% of the time. This is not surprising as the weight matrices in the feed-forward layers of dense models are usually 2x larger than attention layers \cite{kaplan2020scaling} owing to it large hidden size.

    \textbf{Finding \#1 (b)} Interestingly for precision that is supported by the hardware such as FP16/FP32/INT8, prefill latency is significantly faster for higher bit precision compared to unsupported lower bit precisions such as 1/2 bit (\autoref{fig:nr}a).

\end{block}

\begin{figure}[t]
    \centering

    \includegraphics[width=\linewidth]{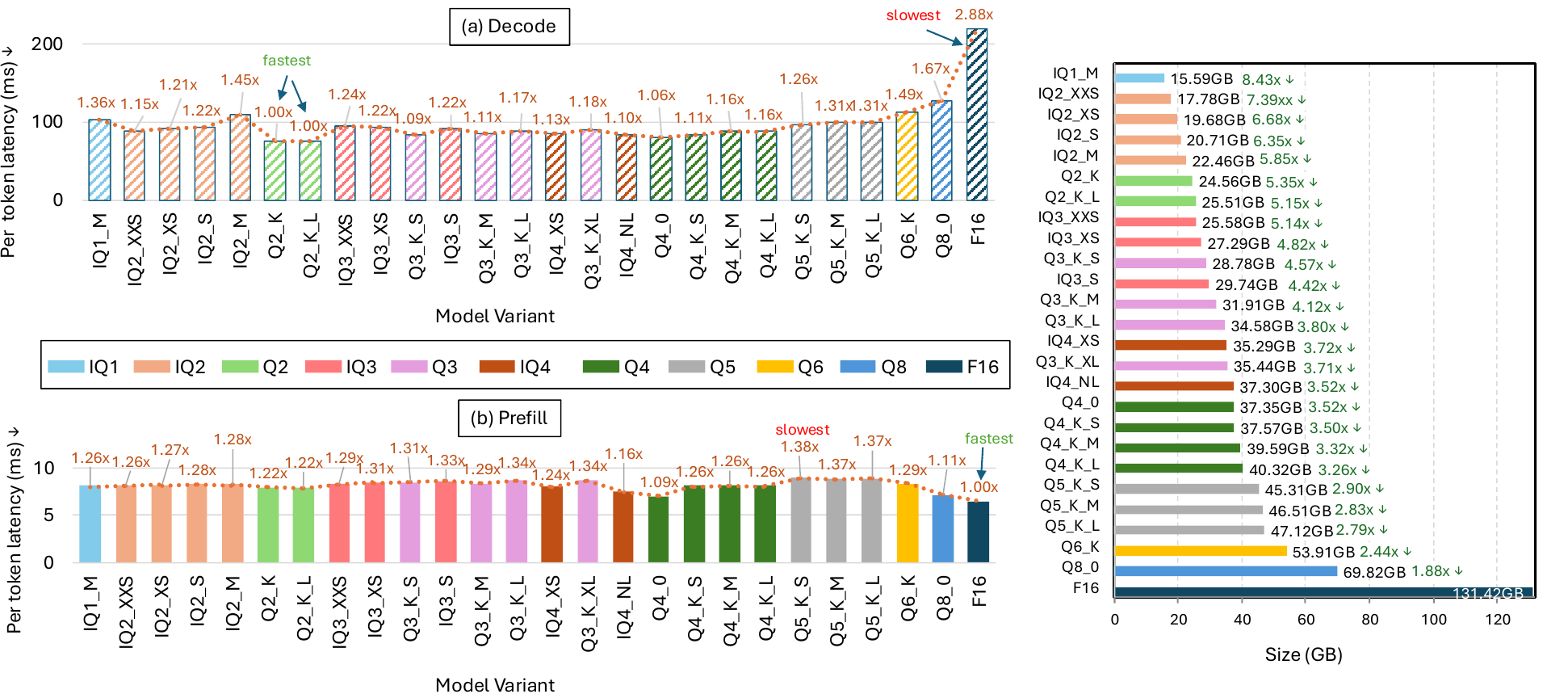}
    \caption{
    Per token inference latency (in ms) of  several Llama 70B model variants as measured on the M2 Ultra for context length 2048 and token generation length of 4096, grouped by different bit precision and sorted by model size. The multiplier on top shows the increase in latency w.r.t the lowest value. \textbf{The non-monotonic nature of the curve indicates that lower bits does not imply faster inference} - despite being 2.09x smaller than Q5\_K\_L, IQ2\_M has higher latency. The rightmost figure highlights reduction in size compared to FP16 variant.}
    \label{fig:inference_latency}
    
\end{figure}
\begin{figure}[t]
    \centering
    \includegraphics[width=\linewidth]{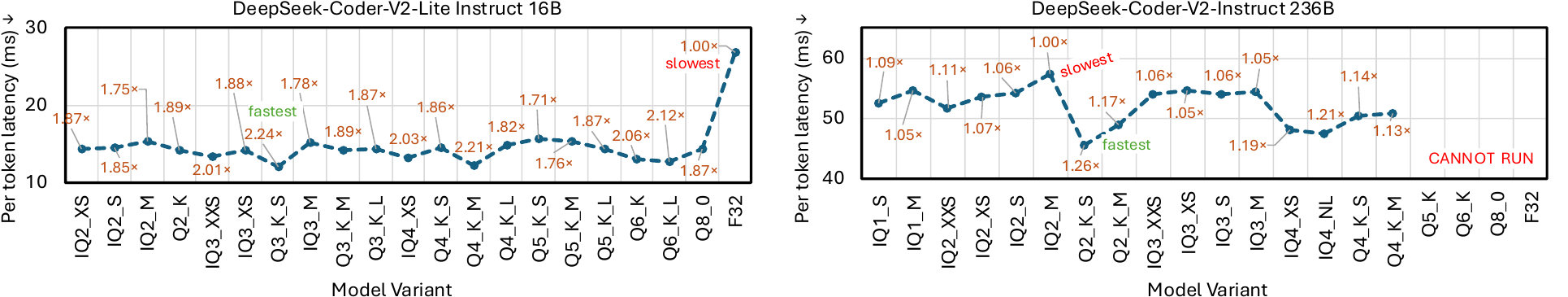}
    \caption{Per token latency (in ms) of (Left) DeepSeek-Coder-V2-Lite Instruct 16B and (Right) DeepSeek-Coder-V2-Instruct 236B on the M2 Ultra for token generation length of 4096.
    Values on top indicate how much faster that variant is compared to the slowest running model. 
    }
    \label{fig:deepseek-latency}
\end{figure}
\subsection{End to end Latency}
\label{sec:e2e}

\autoref{fig:inference_latency} illustrates the prefill and decode latency per token in miliseconds during one inference run of Llama 70B across 26 quantized variants as measured on the \mTwoUltra. We analyze prefill and decode latency separately as their core kernel operations are different. 

Decode is sequential and lacks parallelism \cite{verma2023mastering} - model weights must be reloaded for every token generated; the process is constrained by the available memory bandwidth of the hardware.
The benefits of low bit representation is that the amount of data that needs to be moved reduces, thus faster memory reads/writes can happen - transferring 8-bit data requires 4x less bandwidth than 32-bit. It also means more operations per clock cycle. But this does not equate to a monotonic relation with time taken to generate each token as evident from the non-native bit precisions in \autoref{fig:inference_latency}. Sub-byte model precisions such as the 2.62 bpw IQ2\_M exhibit latency nearly equivalent to that of the 6.56 bpw Q6\_K.

Due to differences in layer architecture and token processing flow, the per-token latency of Deepseek models reported in \autoref{fig:deepseek-latency} is lower than that of dense models at the same parameter scale. While the latency ranking varies slightly, the core finding remains consistent. In Deepseek 236B, the 2-bit IQ2\_M is the slowest, processing at 57.33 ms per token. The 4-bit variant, IQ4
\_NL that is 1.73x the size of IQ2\_M runs 1.21x faster.

FP16 Llama 70B exhibits the lowest latency in prefill. This can seem counterintuitive - higher-precision models like FP16 or INT8 are typically larger in memory and more computationally intensive than quantized variants. Due to the nature of prefill, model weights are reused across all tokens, enabling efficient batch-level parallelism, thus memory bandwidth is not an issue. Native FP16 has no dequantization overhead and INT8 fits well into SIMD lanes ensuring efficient utilization thus reducing latency.
Detailed analysis of these factors are in \autoref{sec:eb}.





Only the M2 Ultra is capable of running Llama 405B - 1-bit (IQ1\_M) variant can run at 442.47 ms per token, 2-bit (Q2\_K) can run for shorter generation length at 1.6s per token as the M2 Ultra memory starts reaching peak saturation, slowing down performance. 

\begin{block}{}
\label{block:2}
\textbf{Common Belief}: Model compression ensures faster inference.

\textbf{Finding \#2}: Lower bit precision does not imply lower latency across all hardware platforms.

From both \autoref{fig:inference_latency} and \autoref{fig:deepseek-latency} we see that Llama 70B and Deepseek 236B models quantized at 2.625 bpw (Q2\_K) is faster than its 1.75 bpw (IQ1\_M) variant irrespective of prefill or decode. For Llama 70B on M2 Ultra in \autoref{fig:inference_latency}a, IQ1\_M at 15.59 GB is 1.36x slower than Q2\_K which is 1.58x larger. At around the same bpw ($\sim$ 2.7), IQ2\_M is 1.45x slower than Q2\_K (9\% larger than the former). 
\end{block}



\subsection{Latency Comparison}
\label{sec:comparison}
\begin{figure}[t]
    \centering
    \includegraphics[width=0.7\textwidth]{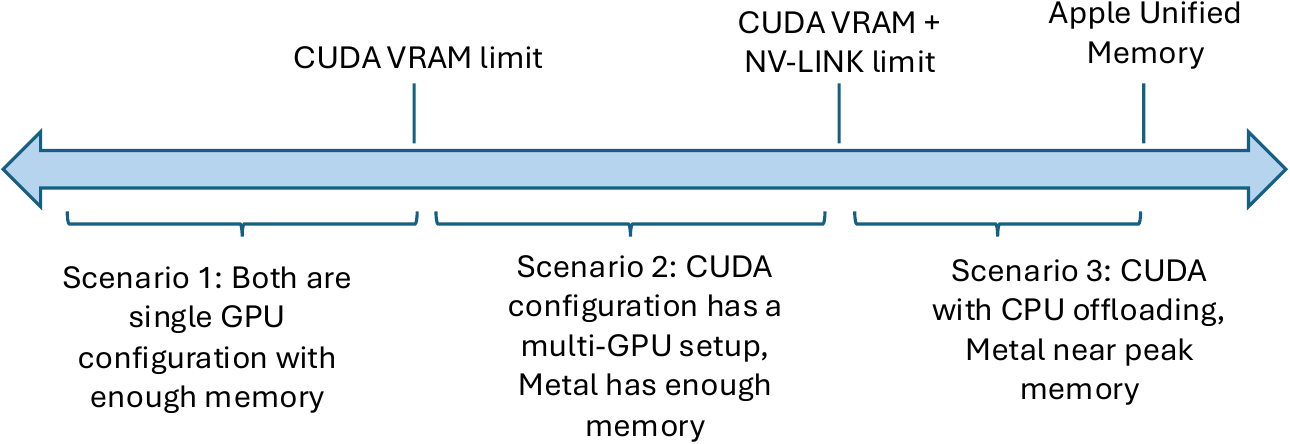}
    \caption{Evaluation scenario.}
    \label{fig:eval_sc}
\end{figure}

\begin{figure}[t]
    \centering
    \includegraphics[width=1\linewidth]{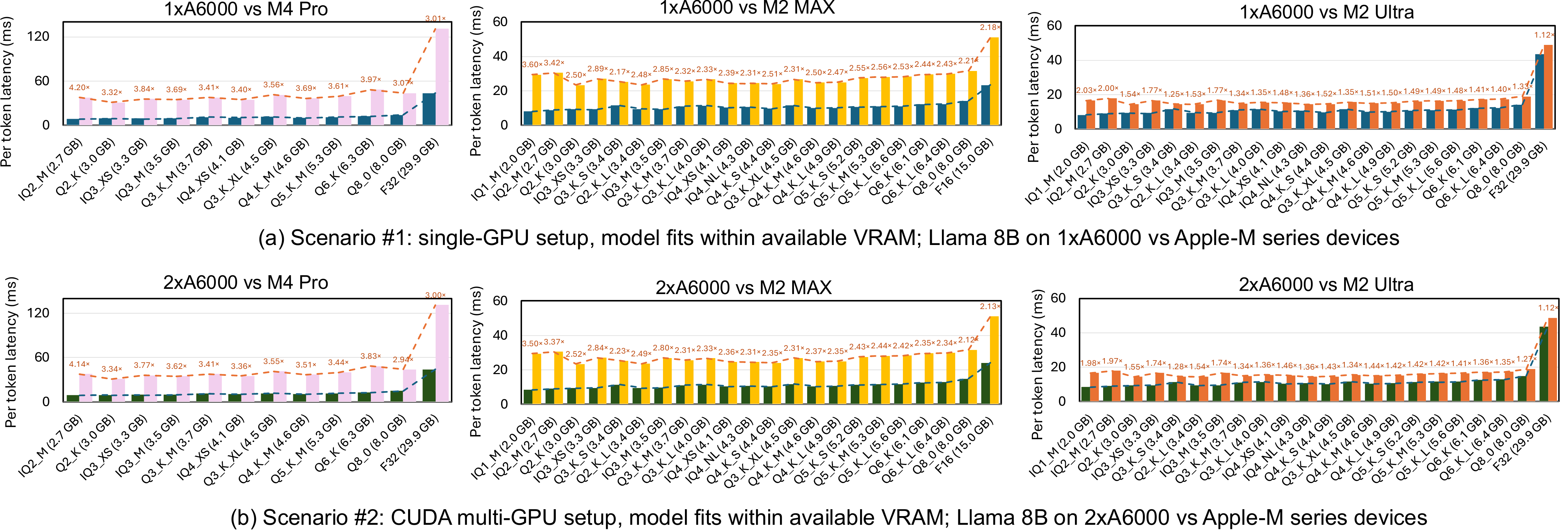}

    \caption{Comparison of inference latency of Llama 8B on NVIDIA GPUs vs Apple Silicon. 
    The multiplier values on top indicate how much slower Metal GPUs are compared to CUDA. Notably, inference on CUDA aligns with the expected trend: lower bit quantization leads to reduced latency. This pattern is highlighted by the larger divergence in the trend lines for the IQ quants across both GPU types. Overall in scenario \#1 and \#2, M-series GPUs are behind by a factor of 1.1x to 4.2x across various model sizes. Between 1xA6000 and M2 Max which are on the same page in terms of effective VRAM, M2 Max largely falls behind in latency, A6000 has 2.1x-3.5x less latency on average. The difference is less pronounced on precisions natively supported by Apple Silicon.}
    
    \label{fig:A6000-vs}
\end{figure}
We compare Apple's Metal to NVIDIA CUDA as their raw compute power allows without considering specific techniques such as Flash Attention \cite{dao2022flashattention}, Apple's deep learning framework CoreML \cite{core-ml} that utilizes ANE or any algorithmic strategy such as speculative decoding \cite{leviathan2022fast}.  
We consider three cases as illustrated in \autoref{fig:eval_sc} - 

\textbf{(1) Single GPU configuration, model fits within available VRAM}: 
\autoref{fig:A6000-vs}a 
portrays the inference latency of Llama 8B on a single A6000 against three M-series GPUs. Both 1xA6000 and M4 Pro feature 48 GB of memory but their inference latencies differ significantly; the M4 Pro exhibits 3.0x to 4.2x higher latency compared to the 1xA6000. Overall, in this scenario, Apple Silicon provides sub par performance with latency being 2.2x-3.6x and 1.1x-2.0x higher in the M2 Max and M2 Ultra respectively compared to 1xA6000 configuration. 

\textbf{(2) CUDA configuration has a multi GPU setup, model fits within available VRAM}: \autoref{fig:A6000-vs}b shows the latency of Llama 8B on  2xA6000 compared to the three M-series devices. Even with a dual GPU communication overhead, 2xA6000 is faster by 2.1x-3.5x and 1.1x-2.0x in comparison to M2 Max and M2 Ultra respectively across model variants. 

\begin{figure}[t]
  \centering
  \begin{minipage}[t]{0.4\textwidth}
    \centering

    \includegraphics[width=0.8\linewidth]{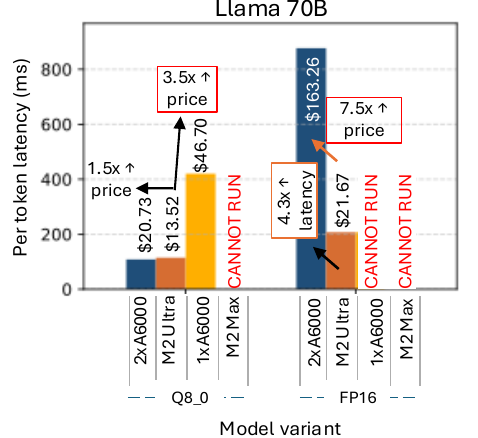}

    \caption{Scenario \#3: Model doesnt fit in CUDA VRAM, Metal near peak memory; 
    Cost/million tokens are shown above each bar.}
     \label{fig:scenario3}
  \end{minipage}\hfill
  \begin{minipage}[t]{0.58\textwidth}
    \centering
    \includegraphics[width=\linewidth]{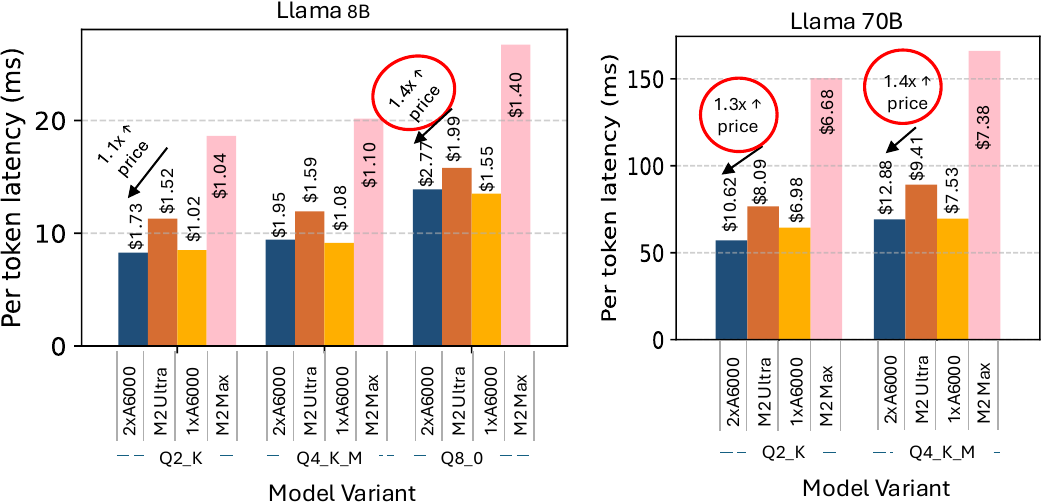}

    \caption{Apple Silicon becomes more cost effective as model size increases.}
     \label{fig:cost-analysis}
  \end{minipage}
\end{figure}

\textbf{(3) Model doesnt fit in CUDA VRAM, Metal near peak memory}: \autoref{fig:scenario3} illustrates the per token latency of Llama 70B as measured on 1xA6000, 2xA6000, M2 Max and M2 Ultra. Llama 70B in FP16 precision at 131.42 GB exceeds the VRAM of 2xA6000 and falls back to CPU for inference, latency significantly increases by 4.3x compared to M2 Ultra; M2 Ultra generates 3.67x more tokens than 1xA6000. 




\subsection{Cost Efficiency}
\label{sec:ce}
\begin{figure}[t]
  \centering
  \begin{minipage}[t]{0.3\textwidth}
    \centering
    \includegraphics[width=\linewidth]{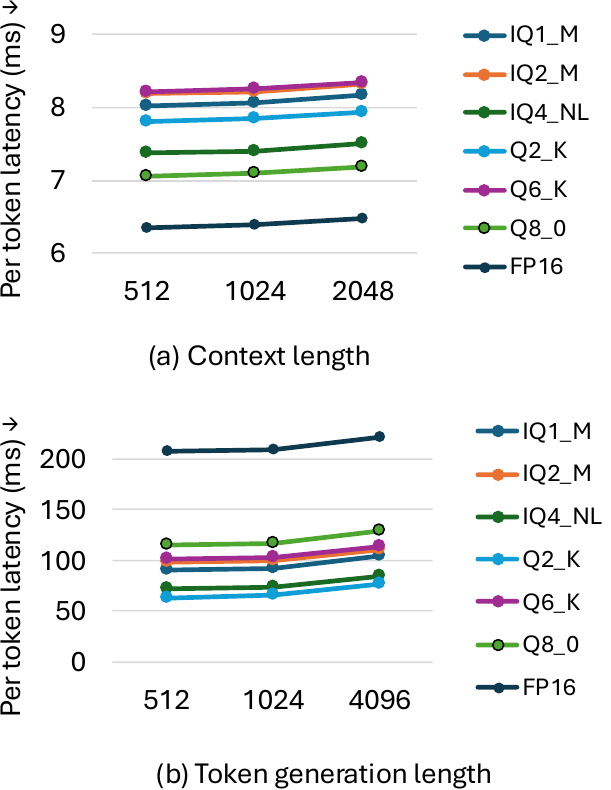}
    \caption{Per token latency (in ms) of selective model variants running on M2 Ultra for varying (a) context length and (b) token generation length.
    }
    \label{fig:pp}
  \end{minipage}\hfill
  \begin{minipage}[t]{0.7\textwidth}
    \centering
    \includegraphics[width=0.95\linewidth]{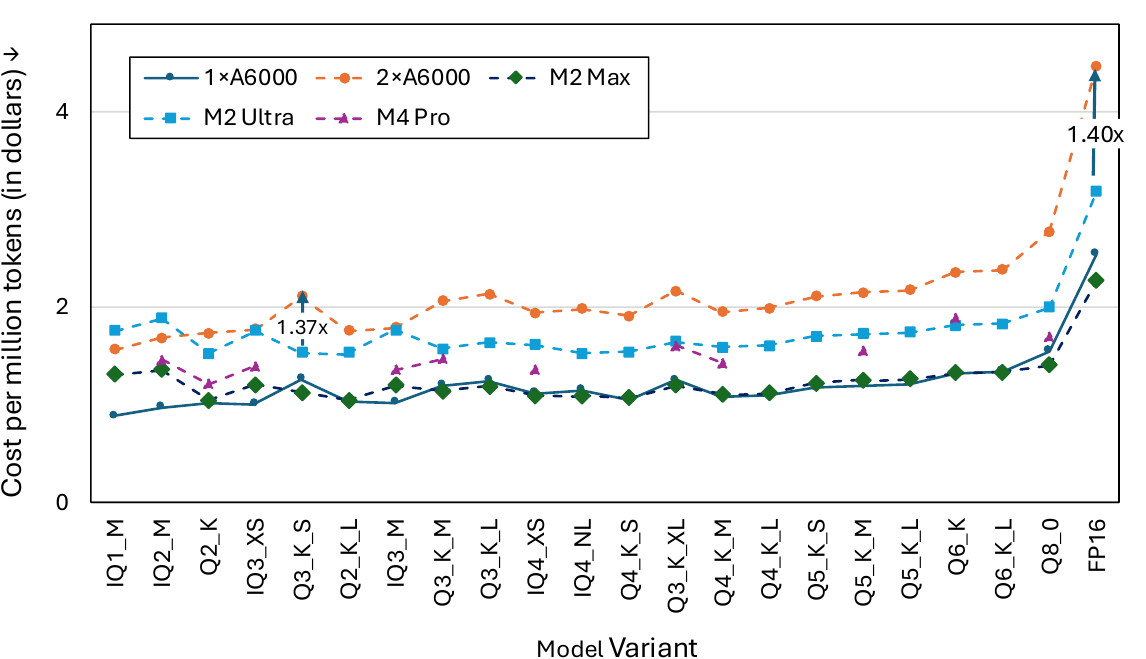}
    \caption{Cost per million tokens for Llama 8B on our hardware configurations.}
    
    \label{fig:cost-per-million}
  \end{minipage}
\end{figure}

\autoref{fig:cost-per-million} shows the cost per million tokens for Llama 8B inference across our evaluation hardware.  The M2 Max demonstrates cost characteristics comparable to 1xA6000, while the M2 Ultra aligns more closely with the 2xA6000 setup.
Under scenario \#1, 2xA6000 incurs upto a 1.58x higher cost compared to the M2 Ultra, for full precision Llama 8B.

Under Scenario \#2, the cost per million tokens for 2xA6000 is consistently 1.2x to 1.6x higher than that of the M2 Ultra across all precisions, excluding the extremely low-bit IQ variants. Specifically, for Llama 8B, the cost ranges from \$1.90 to \$8.13 per million tokens on the 2xA6000, while the M2 Ultra ranges from \$1.75 to \$5.15 per million tokens. M2 Ultra is also capable of generating 3.67x more tokens than 1xA6000 in \autoref{fig:scenario3} while simultaneously being 3.5x more cost effective. 

\begin{block}{}
\label{block:3}
    \textbf{Finding \#3}: Overall, Apple Silicon workstations (M2 Max \& M2 Ultra) are consistently cost-efficient in terms of cost per million tokens, with some exceptions. 
    For smaller models such as Llama 8B, Apple Silicon offers cost efficiency on par with, or superior to, CUDA GPUs but lags behind in performance. However, as model size increases, this performance gap narrows making Apple Silicon efficient and even more cost effective than CUDA GPUs. Relative to the M2 Ultra, the cost per million tokens rises from \textbf{1.4x to 7.5x} when transitioning from a 4-bit to an 16-bit model on the 2xA6000 (ref. Figure \ref{fig:scenario3},\ref{fig:cost-analysis}).
    Except for IQ1\_M and IQ2\_M, the M2 Max and 1xA6000 exhibit comparable cost-efficiency. In contrast, the M4 Pro demonstrates substantially lower performance and increased cost. 

\end{block}





\vspace{-10pt}

\begin{table}[t]
    \centering
    \includegraphics[width=\linewidth]{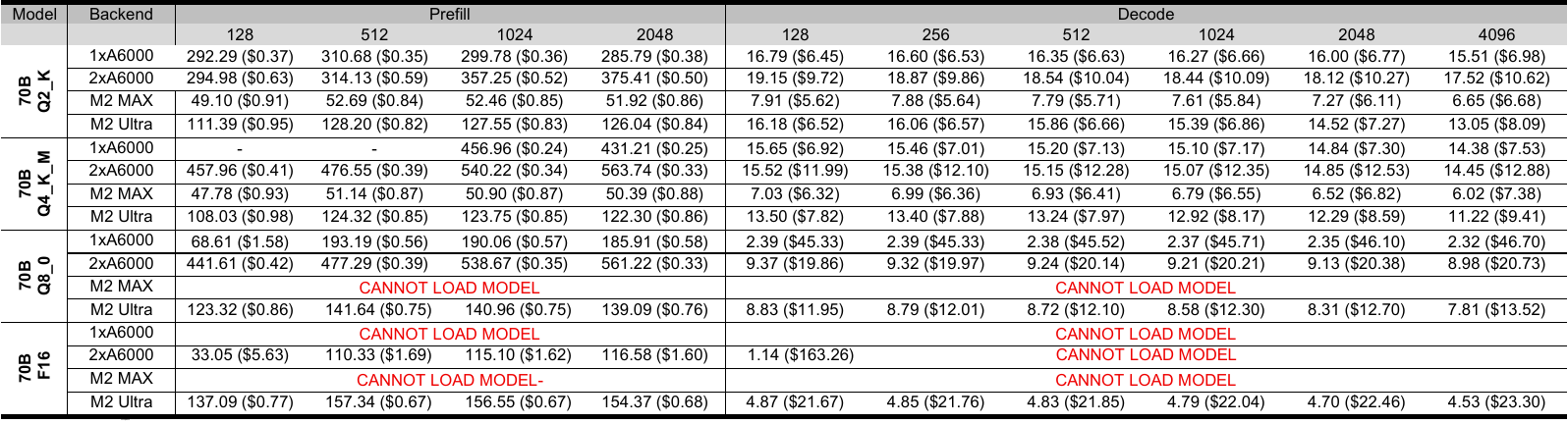}
    \caption{Comparison of cost effectiveness of Apple Silicon vs CUDA over different context and token generation length for Llama 70B. 
    Reported values are tokens per second (TPS) with cost per million tokens inside brackets. Cost generally rises with longer token generation length. Prefill cost efficiency differs across hardware.
    }
    
    \label{tab:perf}
\end{table}

\subsection{Effect of hyperparameter}
\autoref{tab:perf} shows the change in TPS and cost per million tokens with respect to context length and token generation length. 

\textbf{Effect of context length}
\autoref{fig:pp}a plots the time-to-first-token against context length. The curve is increasing for longer prompts (>512) as each token attends to more context.
Time to first token increases with increase in context length on the M2 Ultra. Prefill throughput in CUDA increases upto a certain threshold before declining again.
\textbf{Effect of KV-cache size}: \autoref{fig:pp}b plots the latency increase with increase in KV cache size. As the KV cache size grows, it adds extra latency; matrix-vector operations over growing KV buffers incur more memory reads. Increase in per token latency is proportional to length of generated tokens.


\subsection{Summary of Runtime Cost}
In summary, our fine-grained runtime cost analysis at several bit precision answers RQ1 and RQ2:

(1) \textbf{Is end to end latency related to model size on Apple Silicon?}: We debunk the long-held conception that compressing the model weights must increase its performance speed. We empirically validate that inference latency is not proportional to model precision, rather depends on the quantization scheme used for compression.

(2) \textbf{Is Apple Silicon cost effective in comparison to CUDA for on-device inference?} 
Based on our comprehensive comparison across three distinct scenarios, Apple workstations (M2 Max and M2 Ultra) prove to be the optimal choice and is increasingly cost-effective for larger parameter models (\autoref{fig:scenario3}). The cost gap further widens with model scale on CUDA GPUs, as seen in the higher cost when transitioning from Llama 8B to 70B in \S\ref{sec:ce}. An M2 Ultra rivals a 2xA6000 setup, making Apple Silicon competitive for ultra large language models (>=70B) at higher bit precisions.

(3) \textbf{What is the quantifiable benefit of Apple Silicon’s unified memory?} Running ultra-large language models with
acceptable latency is impossible on contemporary consumer GPUs (NVIDIA, AMD etc.) Apple Silicon offers a single machine workstation without the hassle of configuring a heterogeneous device. Extremely large parameter models such as Llama 405B can run at 1 and 2 bit precision on the M2 Ultra but cannot run on any of our CUDA configurations owing to lack of memory; adding more GPU cards will keep piling on per token cost.

\section{Execution Bottleneck}
\label{sec:eb}
We attempt to answer RQ3 and profile a subset of the available quantization schemes of Llama 70B to understand the hardware performance counters at different model precisions. We compare across different schemes as evidence of
how different kernels can affect low level performance counters which in turn affects latency.


\subsection{Instrumentation}

\begin{figure}[ht]
    \centering
    \includegraphics[width=0.9\linewidth]{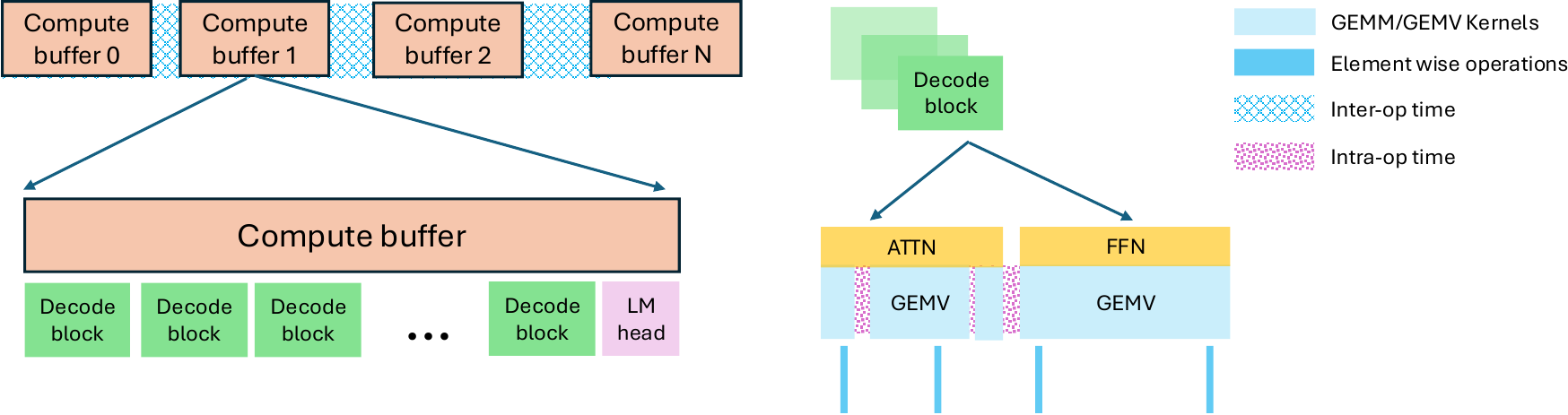}
    \caption{A typical inference execution pipeline.}
    \label{fig:cb}
\end{figure}
We take a top down approach in profiling the GPU system trace. At runtime each token generation is assigned to 1 or 2 compute buffers; one compute buffer usually generates a single token. Multiple kernels can execute in parallel if there is no dependency such as the element-wise operations and GEMM/GEMV kernels  in \autoref{fig:cb}.
The inter operation time is where the compute kernels are awaiting inputs from memory (that has already been executed by the ALU) and intra operation time is mainly the kernel dispatch time. We measure efficiency in terms of generation speed.
The performance of LLM inference on GPUs is intricately tied to the low level hardware features specific to that GPU. Features such as device memory bandwidth, compute utilization, peak FLOP count play an intertwined role in determining the model's holistic performance. 

\subsection{Operator granularity}
In autoregressive decoding, tokens are generated one at a time sequentially; the computation simplifies to a matrix-vector (GEMV) operation in decode phase, unlike the matrix-matrix (GEMM) operations during the prefill stage that runs at higher arithmetic intensity. 
To perform operations on quantized weight matrices, it first need to be dequantized - to make operations more efficient, this dequantization is done on the fly. Inside a quantized kernel such as mul\_mv\_iq1m\_f32 - is a fused dequantization + matrix multiplication operation (ref. \autoref{fig:thruput}c)

\begin{block}{}
\label{block:4}
    \textbf{Finding \#4}: Matrix-Vector operations (GEMV) take up most of the time in decode phase; in prefill Matrix-Matrix operations (GEMM) dominates. \autoref{fig:thruput}c shows the normalized execution time per kernel during prefill and decode, reflecting the mixed-precision distribution of each quantization scheme.
\end{block}


\subsection{Throughput}
\begin{figure}[t]
    \centering
    \includegraphics[width=1\linewidth]{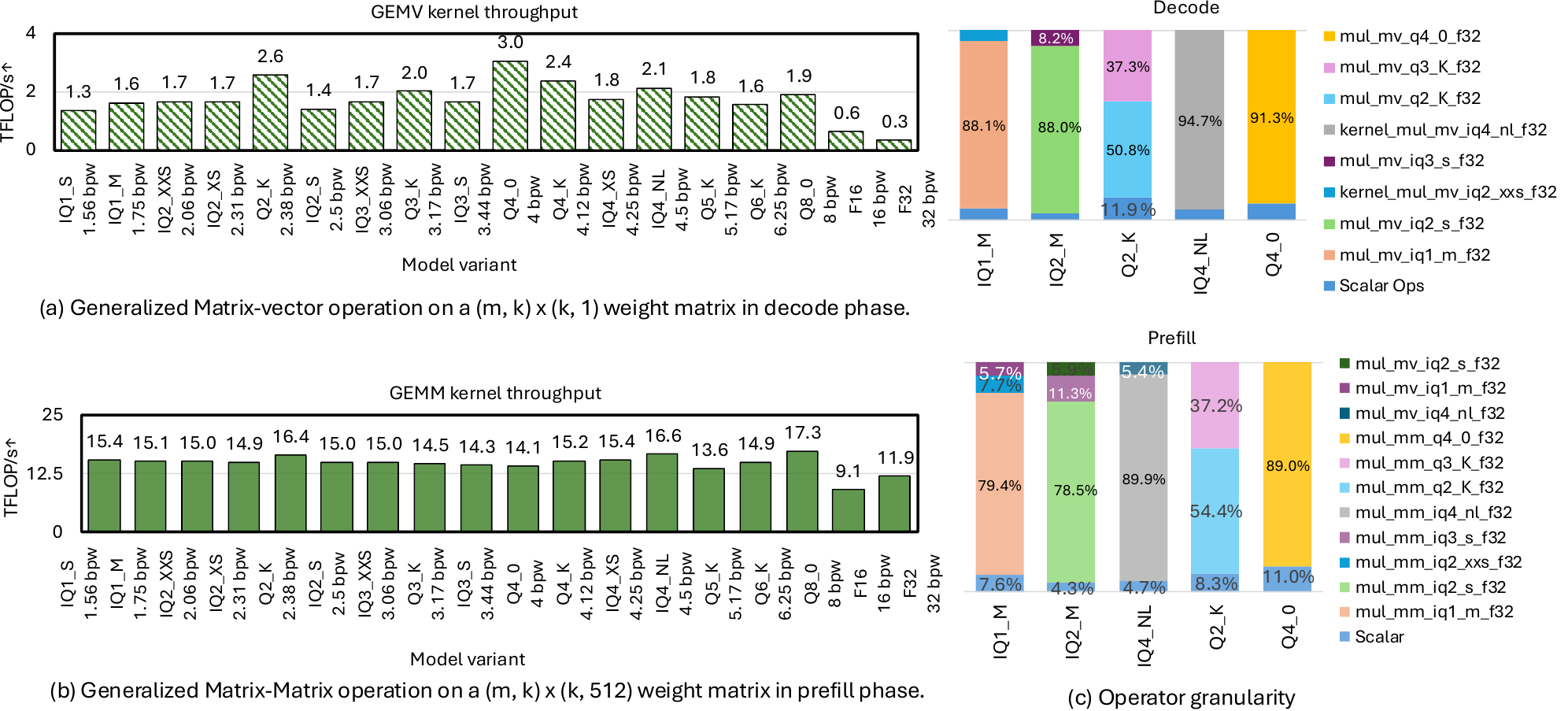}
    \caption{(Left) Throughput of core kernels during prefill (GEMM) and decode (GEMV) as measured on the M2 Ultra for (m, n, k) = (4096, 512, 14336) and (4096, 1, 14336) respectively. (Right) Operator granularity of Llama-70B under different bit precisions.
    }
    \label{fig:thruput}
\end{figure}

In general, higher throughput i.e the number of floating-point operations executed per second is an indicator of good performance. 
Peak throughput and achieved throughput are different as many real-world bottlencks such as memory bandwidth, instruction-level dependencies, kernel overhead etc keeps the achieved throughput well below the theoretical maximum.
\autoref{fig:thruput}a-b illustrate the throughput of several GEMM/GEMV kernels, where each kernel is first dequantized back to FLOAT32 through several steps incuring a 'dequantization overhead' before being multiplied with FLOAT32 activations. 

\begin{block}{}
\label{block:5}
\textbf{Common Belief}: Lower bit precision ensures higher throughput (Ops/s).

    \textbf{Finding \#5 (a)}: Throughput does not scale proportionally with bits per weight (bpw) (ref. \autoref{fig:thruput}a-b); it largely depends on how the quantization scheme is implemented or how much the dequantization overhead is.
    Legacy Q8\_0 has the highest GEMM throughput at 17.37 TFLOPS,
    followed by IQ4\_NL at 16.6 TFLOPS, Q2\_K at 16.37 TFLOPS, and legacy 4-bit at 14.1 TFLOPS.
    
    \textbf{Finding \#5 (b)}: Half-precision throughput on Apple Silicon is lower than full-precision for parallel workloads.
    FP16 and FP32 throughput are at 9.1 TFLOPS and 11.9 TFLOPS respectively. 
\end{block}
\begin{figure}
    \centering
    \includegraphics[width=1\linewidth]{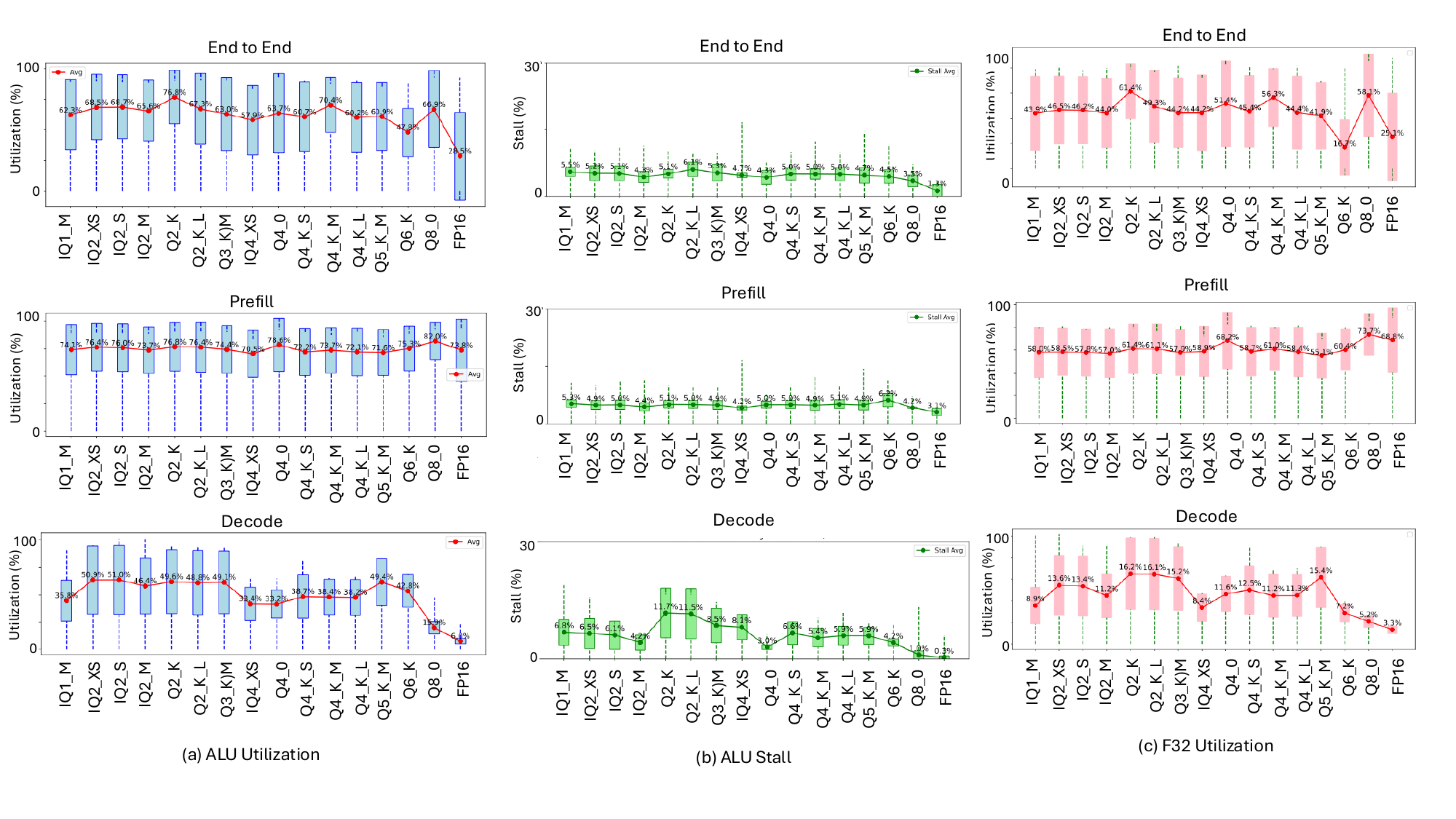}
    \caption{Llama 70B across 14 quantization schemes at context length 64 and token generation length of 10 as measured on the M2 Ultra (a) ALU Utilization (b) ALU stall, measured by the difference of ALU limiter and utilization (c) F32 Utilization}
    \label{fig:pm}
\end{figure}

\begin{figure}
    \centering
    \includegraphics[width=\linewidth]{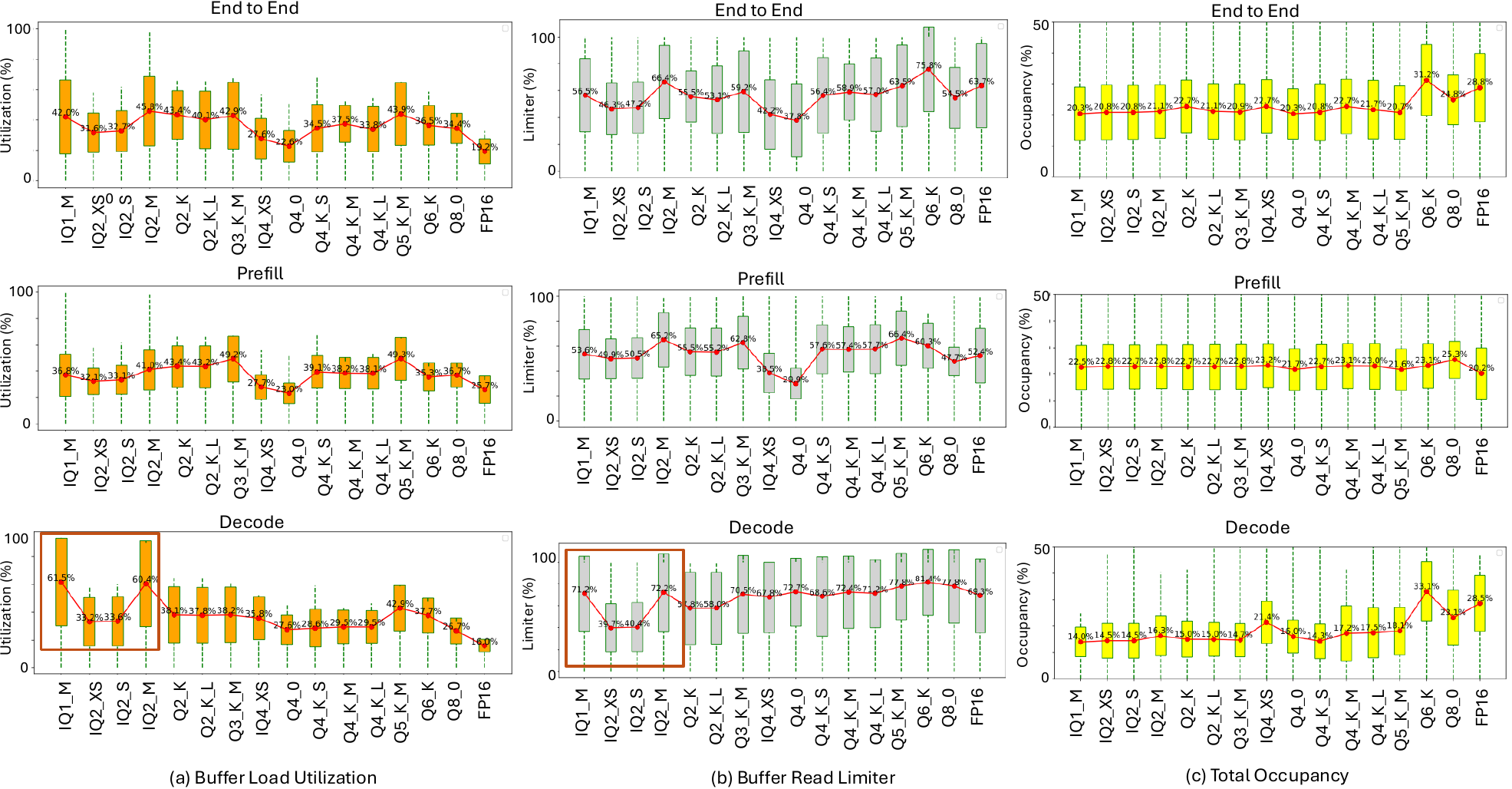}
    \caption{Llama 70B across 14 quantization schemes measured on the M2 Ultra}
    \label{fig:pm2}
\end{figure}

\begin{figure}
    \centering
        \includegraphics[width=0.75\linewidth]{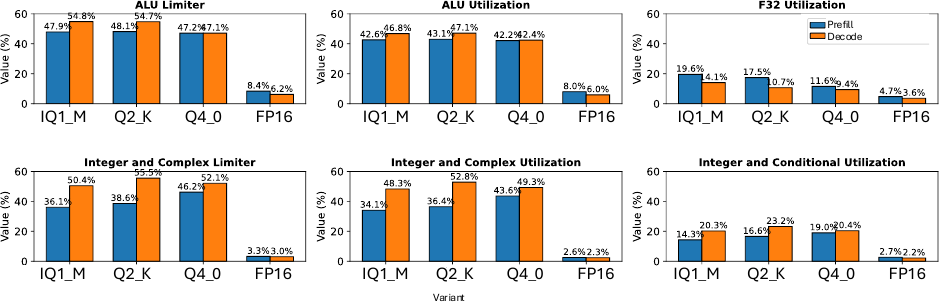}
    \caption{Fine grained profiling of a single transformer block in prefill and decode of 1,2,4 and 16 bit Llama 70B running on M2 Ultra. The results also remain consistent in the Llama 405B decode block.}
    \label{fig:pl}
\end{figure}
\subsection{Performance Counters}
\label{sec:pc}
We look at the GPU performance counters in \autoref{fig:pm} and \autoref{fig:pm2} to gain insights about low-level performance metrics on Apple Silicon. \autoref{fig:pl} profiles a single transformer block for even finer analysis. 
The measured values mostly scale proportionally whether analyzing all blocks or just one.




\subsubsection{ALU utilization} denotes the actual execution of the ALU pipelines. 
\autoref{fig:pm}a reports the ALU utilization and stall of several quantized Llama 70B models. The end to end ALU utilization is the highest in Q2\_K at 76.8\%. 
\textit{ALU limiter} is time during which ALU work is attempted to execute as a percentage of peak ALU performance and is a measure of the amount of work done + stall incurred.
\textit{As bit width increases, ALU utilization tends to decline}—for instance, from 60.7\% at 4-bit to 28.5\% at 16-bit in the end-to-end scenario.
This happens as threads stay busy with memory instructions (load/store) which also leads to fewer stalls, likely because more threadgroups are available.
We observe that the slower quantization schemes also exhibit lower ALU utilization—for example, IQ1\_M shows just 35.8\% ALU utilization during decode, about 15\% lower than the six variants that follow it.
\textit{A notable trend is that higher ALU stalls tend to align with increased ALU utilization}.


\subsubsection{Floating Point utilization} \textit{is almost 5x in prefill than decode} owing to batch parallelism.
\autoref{fig:pm}c reflects both per stage and end to end FP32 utilization of several quantized variants of Llama 70B. 
FP32 utilization remains consistent at 55\%-70\% during prefill. However, during decode differences in utilization makes it apparent as to which variant incurs higher overhead: exhibit IQ1\_M and IQ2\_M at 8.9\% and 11.2\% utilization respectively. FP32 decode utilization declines with bit width (similar to ALU utilization), these deviations across variants suggest additional dequantization overhead.

\subsubsection{INT utilization}
On the contrary, \textit{integer utilization in decode is higher than prefill}. 
Each generated token in decode requires reloading large weight matrices that is dominated by integer arithmetic: address calculation and control flow instructions. 
Both floating point and integer utilization of IQ1\_M is lower than Q2\_K which is understandable as it calls for higher memory (load/store) operations owing to codebook references. 
\autoref{fig:pl} profiles a single transformer block of a few variant from which we see that Q2\_K which is the fastest scheme on M2 Ultra also has the highest INT utilization at 52.8\%.

\subsubsection{Buffer load utilization} measures how often the GPU's memory is actively servicing buffer-load (read) requests. \autoref{fig:pm2}a illustrates an end to end buffer load utilization between 19.2\%-45.8\% across different precisions.
IQ quants with larger codebooks (over 8KB), such as IQ1\_M and IQ2\_M, exhibit notably higher load utilization during decode-60.4\% and 61.5\% respectively-exceeding others by over 30\%.
This indicates larger load transfers per block in IQ quants; unpacking quantized vectors requires frequent load operations beyond the raw tensor operations. Additionally on cache misses, the codebook has to be fetched from the device memory, increasing buffer traffic. 

\subsubsection{Buffer read limiter} reflects the percentage of GPU time that was stalled waiting on buffer reads. A low buffer read limiter means that loads are successfully overlapped with compute and is not stalling the pipeline. A high value means those reads are saturating the memory interface; this is expected to rise as data transfer volume increases with bit precision. As seen in the decode phase of \autoref{fig:pm2}b, IQ1M and IQ2\_M breaks the increasing trendline,  showing higher buffer read limiter values - 71.2\% and 72.2\% respectively - about 30\% more than the two variants between them.
This is explained by IQ1\_M and IQ2\_M transferring significantly more data due to their larger codebooks compared to the two intermediate variants.
\subsubsection{Occupany} \autoref{fig:pm2}c illustrates the total occupancy of several quantized Llama 70B model variants, with end-to-end values ranging from 20.3\% to 31.2\%. 
Higher occupancy reflects more in-flight active SIMD groups, which is common with increasing bit width, as higher bit widths involve more computation per operation and result in longer instruction execution times.
Dispatching more threads will increase the number of SIMD groups in flight but it has to be balanced as then each thread can use fewer registers or less shared memory. 
By default, the benchmark uses all available CPU cores - on the M2 Ultra, dispatching 16 threads at a time yields the best performance.


\subsubsection{Memory bandwidth Utilization}
\begin{figure}[t]
    \centering
    \includegraphics[width=0.9\linewidth]{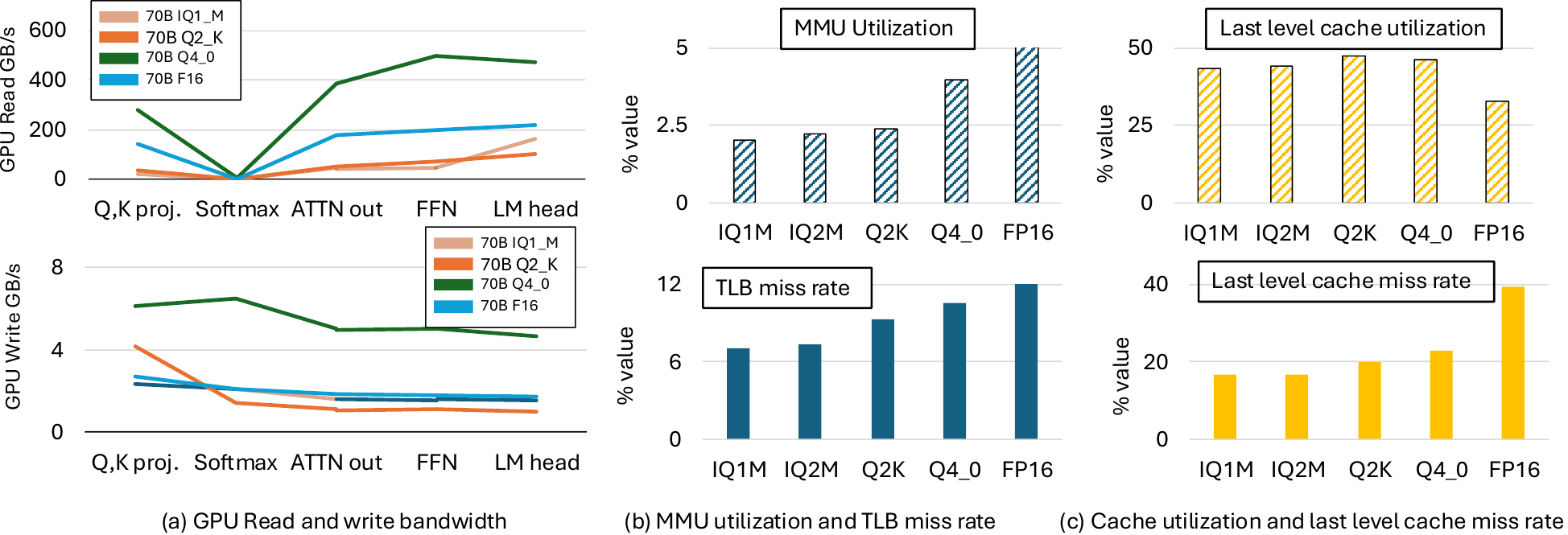}
    \caption{(a) GPU memory bandwidth usage, (b) memory management unit activity, and (c) cache behavior for a single decode block across multiple Llama 70B model variants, measured on the M2 Ultra.}
    \label{fig:mmu-cache}
\end{figure}

About 85\% of the theoretical peak bandwidth is utilized \cite{hubner2025apple}.
As amount of data moved widely varies at different stage in the decode pipeline, we examine a single decode block in \autoref{fig:mmu-cache}a to explore the memory bandwidth usage pattern.
In the typical scenario, the order of bandwidth consumption based on bit precision is IQ1\_M < FP16 < Q2\_K < Q4\_0, with IQ1\_M consistently showing the lowest usage for all key model components. Among the four variants, 
the output head (or LM head) of
Q4\_0 exhibits the highest memory read bandwidth reaching upto 471 GB/s. Since the softmax operation involves minimal data movement, its read bandwidth remains low, around 1–4 GB/s.
FP16 deviates from the typical trend where higher bit-width leads to greater memory utilization, thanks to its direct hardware execution support, lack of dequantization overhead, and reduced overall memory access (no codebook lookup).


\subsubsection{Memory management and caching}
To understand the pattern we examine one single decode block at runtime in \autoref{fig:mmu-cache}.
Last-level cache utilization on the M2 Ultra
reaches nearly 50\% in Q2\_K as seen in \autoref{fig:mmu-cache}c. FP16 has the lowest cache utilization among the five variants.
Memory locality - Cache miss rates range from 17\%–23\%, increasing to 40\% for FP16.
The memory management unit on Apple Silicon becomes less efficient with larger working set sizes, as seen in \autoref{fig:mmu-cache}b; FP16 shows the least spatial locality, with the highest TLB miss rate (13\%) and MMU utilization (6\%).
A lower TLB miss rate indicates better memory locality and performance; however, lower bit precision models, while showing fewer misses, also exhibit lower MMU utilization.

\begin{block}{}
\label{block:6}
    \textbf{Finding \#6}: Apple Silicon favors block based K-quants over codebook based IQ quantization schemes. 
    
    \textbf{Reasoning}: IQ quants are significantly slower on Apple Silicon but is standard on CUDA (ref. \S\ref{sec:comparison}). In block based K-quants, for a block of 16 weights, only one scale and offset value needs to be loaded from memory (DRAM or cache).
    Due to their design, in addition to loading these scale and offset values per block, dequantizing weights to their approximate value in IQ quants requires referencing a table for every 8 or 16 weights which is expensive. This results in increased memory (load/store) instructions.
    The load operation
from a 2 KB to 16 KB table (even from L1 cache) requires more cycles than a simpler ALU operation executed directly on the register, 
e.g. between IQ2\_M that has a codebook of size 16KB and Q2\_K with no codebook, the former is 45\% slower  suggesting that referencing the codebook introduces significant overhead.  
 Analysis of low level GPU counters in \S\ref{sec:pc} reveal discrepancies in execution and memory units reflecting deviations in expected trend lines-certain IQ quants such as IQ1\_M/IQ2\_M consistently have lower compute utilization \& high buffer usage (30\% $\uparrow$ than the rest) owing to a large codebook.
\end{block}
\vspace{-10pt}
\subsection{Analysis}
\label{sec:analysis}
\begin{figure}
    \centering
    \includegraphics[width=\linewidth]{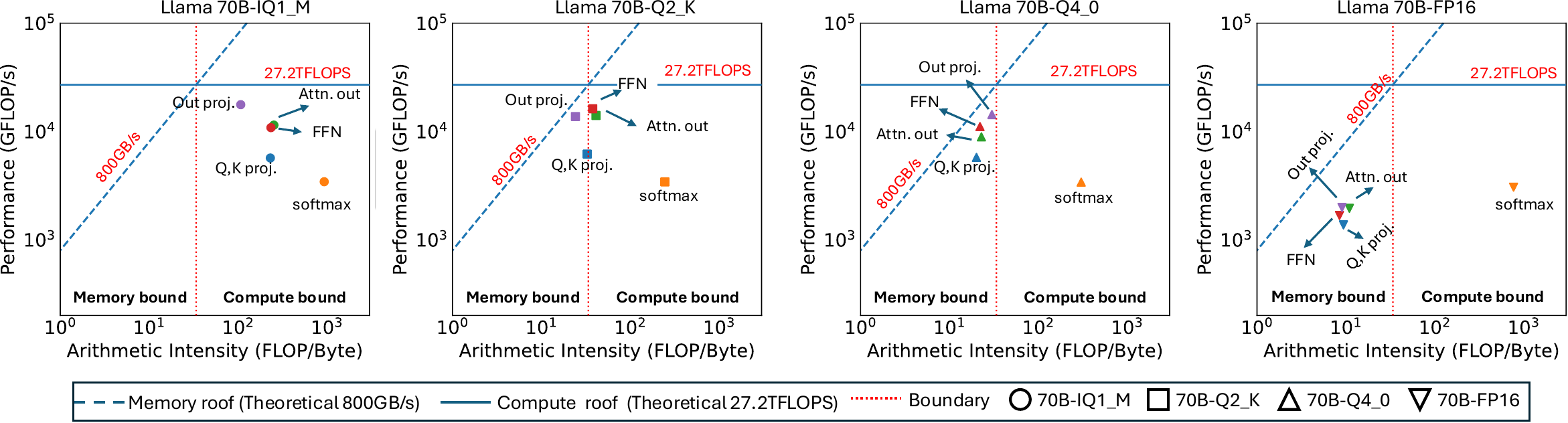}
    \caption{
    Roofline model of a single decode block of Llama 70B on M2 Ultra, detailing the arithmetic intensity of key model components. Softmax operation is always compute bound while others differ - operations in feed forward layers (FFN) and attention output projection (Attn. out) is compute bound at 1\&2 bit-width but memory bound at higher bit precision. The results also remain consistent in Llama 405B decode block.}
    \label{fig:roofline}
\end{figure}

\textbf{Is Compute the Bottleneck or Memory?} Analyzing the roofline model in \autoref{fig:roofline}, we see that within the decode pipeline of Apple Silicon, there is variance in arithmetic intensity per stage. Integer-heavy pointer arithmetic plays a crucial role in determing Arithmetic intensity \cite{ding2019instruction}. Taking into account only simple FP32-FP32 matrix operations is inaccurate. Holitistically the decode phase is bottlenecked by memory and prefill by compute. But Metal GPUs are further bound by arithmetic operations.
There is additional memory lookup at each dequantization step of VQ based quantization schemses. Although it might seem that the additional memory lookup will make them more memory bound, it is on the contrary. The lookups are mostly from cache (L1/L2 or shared memory) so is reused 
and data fetched is insignificant in size (1 or 2 bytes)
thus global memory read access isn’t affected, keeping arithmetic intensity high. 
\begin{block}{}
\label{block:7}
    \textbf{Finding \#7}: IQ quants such as IQ1\_M are bound by compute operations. Because of their complexity, IQ quants require more bit unpacking steps; this bit‑extraction requires multiple scalar bit shift and mask operations. As the weights loaded for 1 bit precision model is smaller ( (i.e memory traffic is very low), arithmetic intensity is higher shifting execution toward the compute‑bound regime (\autoref{fig:roofline}). 
    


\end{block}

\textbf{Dequantization overhead }
When dequantization overhead grows faster than the bandwidth saved, latency can increase even though the tensor is smaller. Thus, for ultra low bit quantization, reducing the dequantization overhead is a primary objective.
Its assumed that loading weights to memory is the bottleneck and cost of dequantization and FP16 computation is small \cite{kim2023squeezellm} but we find for ultra-large language models (Llama 70B) and on Apple hardware, dequantization overhead can be a significant factor. 

\begin{figure}
    \centering
    \includegraphics[width=0.7\linewidth]{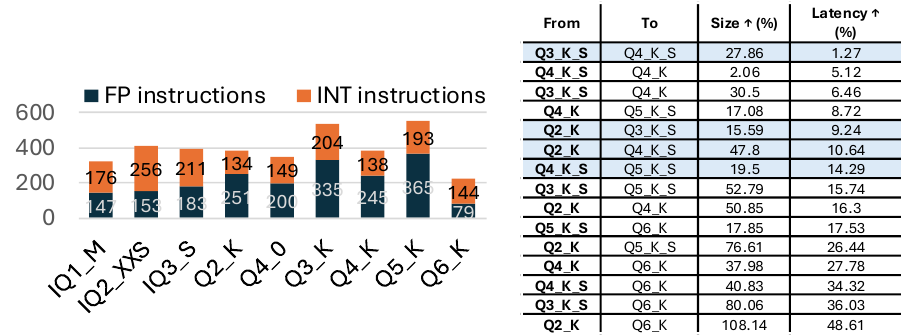}
    \caption{(a) Odd bit-widths (e.g. Q3\_K, Q5\_K) display higher instruction count than the rest portraying the lack of hardware support in Apple Silicon. (b) Penalty incurred as displayed by size and latency increase when converting across bit‑widths.}
    \label{fig:inst_count}
\end{figure}

\begin{block}{}
\label{block:8}
    \textbf{Common Belief}: During decode, loading model weights from memory is the primary bottleneck \cite{kim2023squeezellm}
    
    \textbf{Finding \#8}: The cost of dequantization is significant on Apple Silicon irrespective of quantization schemes.
    
    (1) IQ quants incur delay due to codebook referencing.
(2) All quantization schemes face a dequantization overhead due to bit unpacking, some more than others e.g. Q6\_K kernel executes 1.22x slower than Q8\_0 
(\S\ref{sec:e2e}) despite having lower data traffic (6.56 bpw vs. 8 bpw). This happens as INT8 weights after unpacking fits smoothly into SIMD lanes, whereas unpacking sub-byte bits requires significant bit‑twiddling. Dequantization + scale overhead thus outweighs the bandwidth savings and shows up as a major bottleneck.
\end{block}
\textbf{Why is the dequantization overhead of 3 bit model higher than 1 bit?} The dequantization overhead of IQ1\_M is higher than IQ3\_S owing to more number of operations, more branching as seen from the instruction count of quantized kernels in \autoref{fig:inst_count} and irregular memory access pattern.
A delta correction bit is calculated every 32 weights in IQ1\_M + bit unpacking adds to the overhead.
From analyzing the runtime shader instruction cost we see that IQ quants with higher decode overhead incur a larger proportion of wait time during execution - 26.29\% in IQ1\_M vs 19.24\% in IQ3\_S.


\paragraph{How are bits packed?}
ALU datapaths and register files on modern GPUs are usually 16 bit or  wider (FLOAT16/INT16 256-bit vector lanes). Existing hardware cannot natively store fractions of an INT8; thus to represent irregular bits (1/2/3 etc.) multiple sub-int8 bits representing several values are packed together into one FLOAT16  which is later disassembled with bit-shift operations. Regular bit-widths such as 4-bit fields can fit cleanly into a 16-bit register with zero wasted bits, for others such as 1/3/5/6 bit, the additional bits are stored in separate bytes. 
This pattern is evident in the table in \autoref{fig:inst_count}: moving from Q3\_K\_S to Q4\_K\_S (3-bit to 4-bit) causes only a 1.27\% increase in latency.
However, moving from Q4\_K\_S to Q5\_K\_S (4-bit to 5-bit) results in a sharp 14.29\% increase in latency although increase in size is not drastically different.
Transitions from Q2\_K (2-bit) to Q3\_K\_S and Q4\_K\_S show similar latency increases (9.24\% and 10.64\%), reinforcing that odd widths tend to add more overhead regardless of direction.

\begin{block}{}
\label{block:9}
\textbf{Finding \#9}: Apple Silicon lacks dedicated hardware units for tensor intensive workloads common in modern ML applications, that hinders its performance and amplifies the dequantization overhead. 

\textbf{Reasoning}: Owing to how bits are packed, irregular bit widths such as 3 or 5 bit incur
a higher overhead, needing more bit-shifting, masking and scale operations than regular 4/8-bit alignment. This is evident from the higher instruction count of Q3\_K and Q5\_K in \autoref{fig:inst_count}a.
This misalignment increases processing complexity, as a result, switching from an odd to an even bit-width (e.g. Q3 $\rightarrow$ Q4) tends to have a smaller latency increase, while transitioning from an even to an odd width (e.g., Q4 $\rightarrow$ Q5) shows a much larger jump in latency (ref. \autoref{fig:inst_count}b). 
Even bit precision (e.g 2/4/8 except 6) is favored over odd bit (1/3/5) precision.

\textbf{Why is IQ1\_M slower in M2 Ultra but faster on the A6000?} 
A6000 also does not have support for irregular bitwidths but the dequantization overhead is largely amortized/hidden behind faster compute owing to tensor cores that deliver high INT8/INT4 throughput \cite{wu2023understanding}. M2 Ultra lacks dedicated low‑bit MMA units \cite{msl}; IQ1’s bit extraction executes on generic SIMD integer/FP pipelines, making dequantization the bottleneck. 
\end{block}


\textbf{Implication of Codebook dimension}
Codebook dimensions vary across schemes: IQ1\_M stores 2048 vectorsx8B (16KB), IQ2\_S uses 1024x8B (8KB), IQ2\_XS 512x8B (4KB) and IQ3\_XXS 256x4B (1KB) while the non-linear \_NL quants have only 16 bytes of metadata.
To determine whether the increased latency in IQ quants is driven by codebook lookups or by the higher ALU cycles needed for dequantization, we perform two controlled experiments 
(1) increasing the codebook dimension of IQ3\_XXS from 1KB to 16KB as in IQ1\_M. (2) decreasing the data type of IQ1\_M from UINT32 to UINT8. We observe a somewhat notable increase in speed when decreasing datatype, conversely, there is no change when the codebook dimension is modified - implying that the the dequantization overhead is the primary determining factor contrinuting to latency.













\subsection{Summary of runtime bottlenecks}
We find that several factors such as design of quantization scheme, dequantization overhead, memory bandwidth and model architecture determines the time taken for one forward pass.

(1) \textbf{Throughput}: in Apple Silicon is not proportional to bits per weight (bpw) rather depends on how the quantization scheme is designed. Thus at runtime, inference is bottlenecked by lower throughput.

(2) \textbf{Cost of ALU operation}:
For lack of dedicated cores, ALU operations in Apple Silicon are not fused, thus requires more ALU cycles to execute instructions. Additional memory operations in IQ quants adds to the inference cost.


(3) \textbf{Irregular bits} are more inefficient on Apple Silicon as they require more instructions to unpack and the unpacked bits don’t align with the 8 or 16‑bit SIMD lanes.

(4)\textbf{ Are the kernels well utilized?} Observing the ALU and buffer load utilization, there is scope to better implement the current kernels keeping Apple Silicon in mind.

\section{Recommendation}
\label{sec:recommendation}
\textbf{For ML practitioners:}
Certain quantization schemes are specifically suited for Apple Silicon and can be kept in mind while deploying on-device models. 
(1) At a similar level of memory consumption, K-quants are superior over codebook based IQ quants in Apple Silicon in terms of inference speed, perplexity and cost effectiveness. \autoref{app:A} reports model perplexity.
(2) If a slight drop in accuracy is acceptable, we recommend 2 bpw block based quants (Q2\_K) for Apple devices that cannot fit models at 4 bpw. 
(3) If the model fits in memory, legacy quantization schemes such as Q4\_0, Q8\_0 are superior and should be chosen over irregular
bit widths such as Q3\_K or Q5\_K in Apple Silicon.
(4) We recommend IQ4\_NL to gain a higher accuracy over 2 bpw quants. It maybe worthwhile for the machine learning community to pursue \_NL quants at < 4 bit precision.

\textbf{For hardware vendors:} To address the performance gap in comparison to contemporary GPU vendors, the following can be addressed:
(1) There is a strong case for integrating dedicated cores into Apple Silicon, analogous to tensor cores which would be optimized for large‑scale LLM workloads; existing ANE is limited in capacity and not suited for such workload.
(2) Native support for lower‑precision arithmetic such as INT4/FP8 in Apple Silicon will enable running sub-byte model precisions directly on the hardware. (3) Documentation for low level kernels in Apple Silicon is mostly closed source preventing developers from optimizing kernels at the ISA level. 
(4) Compute budget of Apple Silicon needs to be increased particularly to process the scalar integer ALU instructions faster - this will remove the primary bottleneck for IQ1\_M like quants that are compute bound.
\section{Related work} 
Recent research has investigated leveraging Apple Silicon’s unified memory system and built-in accelerators to support a wide range of computational tasks including simple classification algorithms \cite{struniawski2024exploring}, scientific computing \cite{hubner2025apple, kenyon2022apple} and numerical simulations \cite{gebraad2023seamless}. \cite{tang2025scaling} reports gain in prefill time of their inference framework on M4 Pro.
\cite{li2024large} surveys efficient generative LLM inference across a versatile choice of hardware platforms such as CPU, GPU, FPGA, ASIC and PIM/NDP excluding Apple Silicon.
Serving of quantized large language models is inherently challenging because of the overhead associated with dequantization, which manifests both as a hardware limitation and an algorithmic bottleneck \cite{lin2405qserve}. QServe \cite{lin2405qserve} attempts to mitigate the dequantization overhead in INT4 quantization using register-level parallelism on CUDA GPUs. Several quantization algorithms exist \cite{tseng2024quip, frantar2022gptq, kim2023squeezellm, liu2025vq} with different order of complexity during dequantization. 
Prior studies \cite{egiazarian2024extreme} on quantization schemes have largely concentrated their effects on model perplexity, our work highlights how hardware characteristics influence the efficiency of a diverse set of quantization schemes on Apple Silicon.


\section{Conclusion}
We conduct a comprehensive evaluation of Apple Silicon to understand its hardware characteristics and implications for on-device LLM inference. We benchmark its performance against contemporary NVIDIA GPUs at similar price range and establish its superiority for on-device inference of extremely large language models. 
Our investigation yields several key insights and suggests practical recommendations for both ML practitioners and hardware designers to make hardware aware choice of models at different bit precisions. We explain the performance gap through fine grained, low level profiling of Apple Silicon.

\bibliographystyle{plainnat}
\bibliography{main}

\newpage
\appendix
\section{Appendix}
\label{app:A}
\begin{figure}[h]
    \centering
    \includegraphics[width=0.5\textwidth]{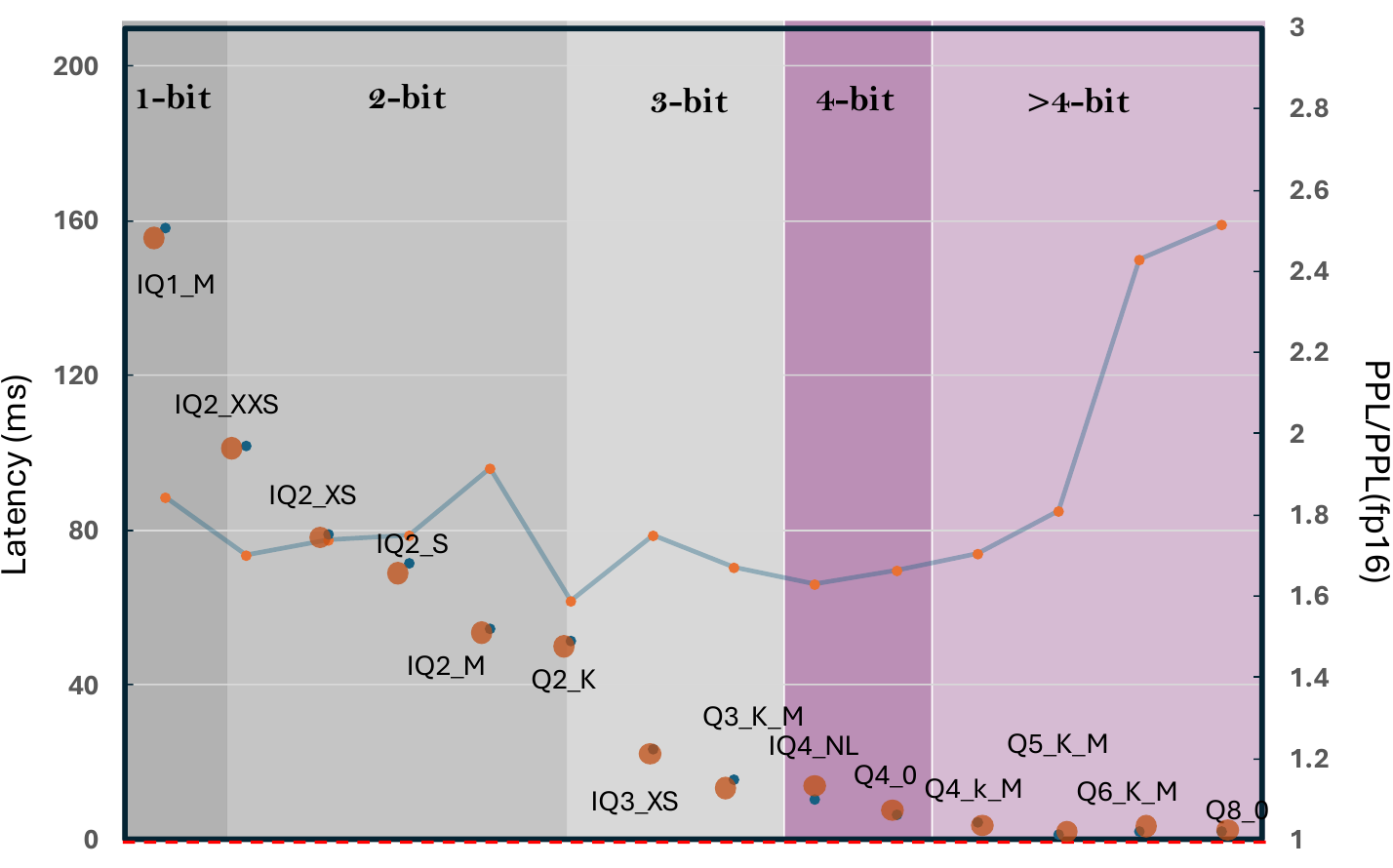}
    \caption{Perplexity of (Llama-70B) on wiki-test-raw dataset. Right axis shows increase in perplexity compared to the f16 variant and left axis shows the per token latency in miliseconds. <4 bit models have higher perplexity. Perplexity of 2 bit IQ2\_M and Q2\_K although similar, Q2\_K is 1.45x times faster than IQ2\_M. \textbf{Perplexity is solely based on bit-width.}}
    \label{fig:perplexity}
\end{figure}

\end{document}